\documentclass[aps,prb,reprint,twocolumn,longbibliography,superscriptaddress]{revtex4-2}

\usepackage{amsmath}
\usepackage{amssymb}
\usepackage{mathrsfs}
\usepackage{empheq}
\usepackage{etoolbox}
\usepackage{graphicx}
\usepackage{color}
\usepackage{bm}
\usepackage{epstopdf}
\usepackage{mathtools}
\usepackage{soul}

\usepackage[dvipsnames]{xcolor}
\usepackage[colorlinks=true]{hyperref}
\hypersetup{
    colorlinks=true,
    linkcolor=blue,
    citecolor=blue,
    urlcolor=blue
}

\newcommand{\ep}{\varepsilon}

\newcommand{\be}{\begin{equation}}
\newcommand{\ee}{\end{equation}}
\def\ba{\begin{aligned}}
\def\ea{\end{aligned}}
\newcommand{\bea}{\begin{eqnarray}}
\newcommand{\eea}{\end{eqnarray}}

\renewcommand{\Re}{{\rm \, Re\,}}
\renewcommand{\Im}{{\rm \, Im\,}}

\renewcommand{\hat}[1]{{\widehat #1}}

\makeatletter
\DeclareFontFamily{OMX}{MnSymbolE}{}
\DeclareSymbolFont{MnLargeSymbols}{OMX}{MnSymbolE}{m}{n}
\SetSymbolFont{MnLargeSymbols}{bold}{OMX}{MnSymbolE}{b}{n}
\DeclareFontShape{OMX}{MnSymbolE}{m}{n}{
    <-6>  MnSymbolE5
   <6-7>  MnSymbolE6
   <7-8>  MnSymbolE7
   <8-9>  MnSymbolE8
   <9-10> MnSymbolE9
  <10-12> MnSymbolE10
  <12->   MnSymbolE12
}{}
\DeclareFontShape{OMX}{MnSymbolE}{b}{n}{
    <-6>  MnSymbolE-Bold5
   <6-7>  MnSymbolE-Bold6
   <7-8>  MnSymbolE-Bold7
   <8-9>  MnSymbolE-Bold8
   <9-10> MnSymbolE-Bold9
  <10-12> MnSymbolE-Bold10
  <12->   MnSymbolE-Bold12
}{}

\let\llangle\@undefined
\let\rrangle\@undefined
\DeclareMathDelimiter{\llangle}{\mathopen}%
                     {MnLargeSymbols}{'164}{MnLargeSymbols}{'164}
\DeclareMathDelimiter{\rrangle}{\mathclose}%
                     {MnLargeSymbols}{'171}{MnLargeSymbols}{'171}
\makeatother

\begin{document}

\title{Entropy and de Haas-van Alphen oscillations of a three-dimensional \\ marginal Fermi liquid} 

	\author{P. A. Nosov}
	
	\affiliation{Stanford Institute for Theoretical Physics, Stanford University, Stanford, California 94305, USA}
 \affiliation{Kavli Institute for Theoretical Physics, University of California, Santa Barbara, CA 93106, USA}
	
	\author{Yi-Ming Wu}
	\affiliation{Stanford Institute for Theoretical Physics, Stanford University, Stanford, California 94305, USA}
	
	\author{S. Raghu}
	
	\affiliation{Stanford Institute for Theoretical Physics, Stanford University, Stanford, California 94305, USA}
	
	\affiliation{\hbox{Stanford Institute for Materials and Energy Sciences, SLAC National Accelerator Laboratory, Menlo Park, CA 94025, USA}}
	\date{\today}
	
	\begin{abstract}
	We study de Haas-van Alphen oscillations in a marginal Fermi liquid resulting from a three-dimensional metal tuned to a quantum-critical point (QCP). 
We show that the conventional approach based on extensions of the Lifshitz-Kosevich formula for the oscillation amplitudes
becomes inapplicable when the correlation length exceeds the cyclotron radius. This breakdown is due to
(i) non-analytic finite-temperature contributions to the fermion self-energy (ii) an enhancement of the oscillatory part of the self-energy by quantum fluctuations, and (iii) non-trivial dynamical scaling laws associated with the quantum critical point. We properly incorporate these effects within the Luttinger-Ward-Eliashberg framework for the thermodynamic potential by treating the fermionic and bosonic contributions on equal footing. As a result, we obtain the modified expressions for the oscillations of entropy and magnetization that remain valid in the non-Fermi liquid regime.
	\end{abstract}

	\maketitle
\section{Introduction}
Magnetic oscillations in metals - the periodic variation of virtually all physical quantities with an external magnetic field - are remarkable macroscopic quantum effects.  Such quantum oscillations are periodic in $1/B$, where $B$ is the external magnetic field, and are routinely used in the experimental characterization of metals \cite{shoenberg_1984,abrikosov2017fundamentals}.  Among the foremost examples is the field-induced oscillation of magnetization, known as the  de Haas-van Alphen (DHVA) effect, which has played a pivotal role in the study of Fermi surfaces of elemental metals, and even those of strongly correlated systems such as the cuprates \cite{vignolle2008quantum,Sebastian2010,QO_Tc_review,barivsic2013universal,Ramshaw2015}, iron-pnictides \cite{Shishido2010,Walmsley2013}, and heavy fermion materials \cite{McCollam2005,Hornung2021}.

A quantitative theory of DHVA oscillations was first put forward by Lifshitz and Kosevich (LK) \cite{lifshitz1956theory}, who  considered the oscillatory behavior of the thermodynamic potential of a clean, three-dimensional gas of non-interacting electrons with
arbitrary dispersion.  According to the LK theory, the oscillation frequencies reveal areas of ``extremal orbits", {\it i.e.}, the Fermi surface orbits having extremal momentum-space areas in the plane perpendicular to {\bf B}.  In addition, from the temperature-dependence of the oscillation amplitude for the $k$-th harmonic (also known as the LK formula)
\begin{equation}
	A_k(T)=\frac{T}{2\sinh(2\pi^2T k/\omega_c)}\label{eq:LK1}
\end{equation}
with $k=1,2,...$,
it is possible to extract cyclotron masses associated with extremal orbits through $\omega_c=eB/m$. In the presence of impurities, the oscillations are exponentially damped by a finite elastic scattering rate.  

Despite the obvious presence of sizable interactions in real metals, nearly all experimental observations of DHVA oscillations to date are interpreted in terms of the LK theory, often with striking success.  The reason for such robustness of the predictions LK theory was given  by Luttinger \cite{LuttingerDHVA}, and  by Bychkov and Gorkov \cite{GorkovBychkov}, who showed that so long as the metal remains a Fermi liquid, many-body interactions do not alter the LK predictions; instead many-body effects are manifest in renormalized parameters, such as altered extremal orbits and the fully dressed cyclotron masses $m^*$ as extracted from the LK formula in Eq.\eqref{eq:LK1}.  We refer to such an  extension of LK theory to interacting systems as the extended LK paradigm,  
 which provides a well-defined theoretical prescription for interpreting DHVA oscillations. In practice, one starts with the expression for the thermodynamic potential, a functional of fully dressed propagators and self-energies. In a magnetic field, all such quantities contain both uniform and oscillatory components. The extended LK prescription involves neglecting both the temperature dependence and oscillatory components of the fermion self-energy.  Doing so, one can show that the oscillations satisfy LK theory albeit with renormalized parameters. Specific applications of this theory include coupled electron-phonon systems \cite{FowlerPrange,Engelsberg1970,Wasserman1981} and disordered two-dimensional Fermi liquids \cite{Maslov2003,Adamov2006}, also see \cite{Wasserman1996} for a review.  

In a Fermi liquid, the extended LK method is well-justified.  For instance, the  Fermi liquid self-energy is quadratic in temperature, which can be neglected.  Moreover, the oscillatory piece of the self-energy is suppressed by a factor $\left(\omega_c/\mu \right)^{3/2}$ compared to the smooth part, where $\mu$ is the Fermi energy. As a consequence, the corresponding contribution to the oscillations of the thermodynamic potential is suppressed by a factor $\left(\omega_c/\mu \right)^{1/2}$ compared to the leading term, and thus, can also be neglected. It is far from clear, however, whether such a prescription remains valid when a metal is tuned to a quantum critical point (QCP), where interactions induced by soft order parameter fluctuations become singular, resulting in the breakdown of Fermi liquid theory, and in non-analytic temperature dependence of all self-energies. 


Indeed, the naive adoption of the extended LK prescription  to the case of a metal with singular interactions was questioned recently\cite{constraints},  from a perspective that it results in the violation of the third law of thermodynamics (Nernst's theorem), which requires that the entropy density and the specific heat must vanish as $T \rightarrow 0$.
A simple manipulation shows that this requirement implies the temperature derivative of the oscillation amplitude, $\partial A_k/\partial T$, must vanish in the zero temperature limit.
For concreteness, we will consider a 3D metal coupled to some collective bosons which can be either undamped or overdamped. When the system is far away from the QCP it behaves like a conventional Fermi liquid, and the application of the LK prescription still leads to Eq.\eqref{eq:LK1} (with renormalized quantities) from which it is obvious to see that $A_k(T)-A_k(T=0){\sim}{-}T^2$ at $T\ll \omega_c$ and therefore $\partial A_k/\partial T=0$ as $T\to0$. However, when the system is close to the QCP, singular fermion interaction mediated by critical bosons results in a marginal Fermi liquid (MFL) behavior. In this regime, working within the LK paradigm leads to $A_k(T)-A_k(T=0)\sim T\ln T$ at low temperatures, and therefore the apparent divergence of $\partial A_k/\partial T$ as $T\to0$, i.e. the violation of the third law.
This observation leads to the breakdown of the LK paradigm, and calls for a more careful analysis of the quantum oscillations for a metal in a quantum critical regime.

\begin{figure}
    \centering
    \includegraphics[width=0.95\columnwidth]{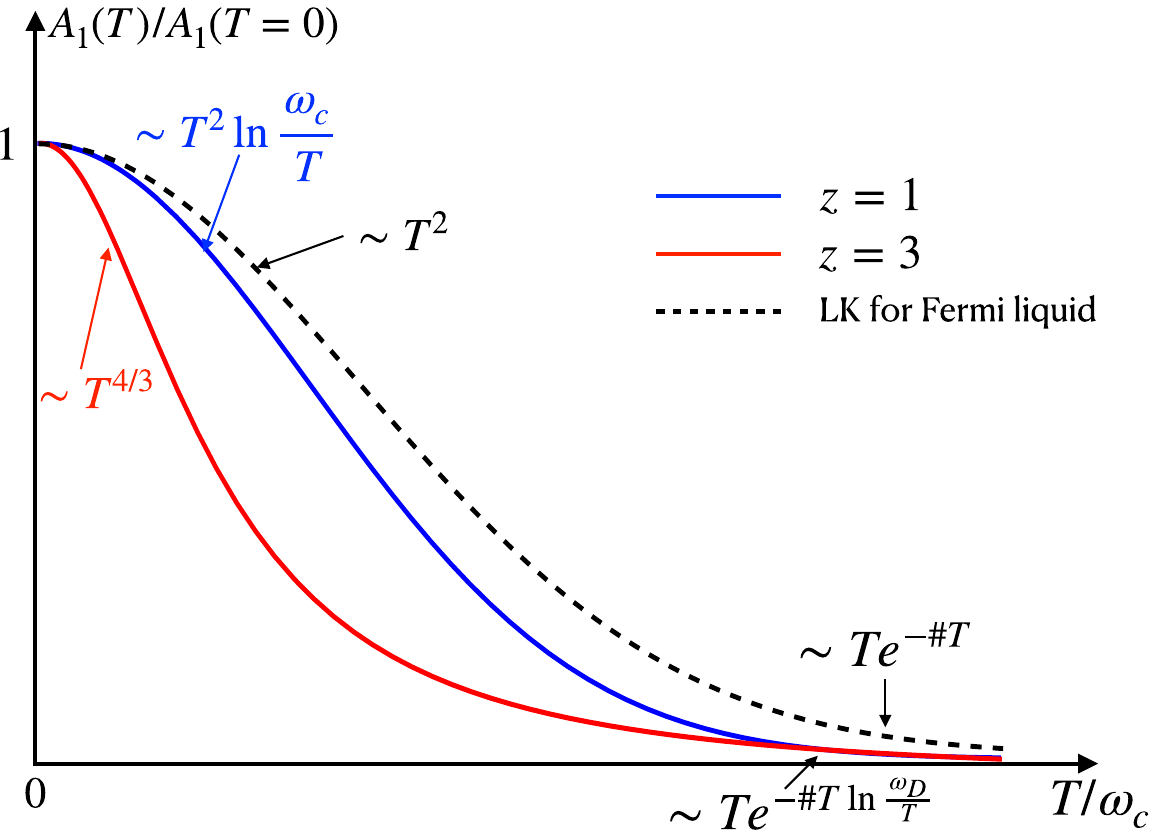}
    \caption{The temperature-dependence of the DHVA oscillation amplitude $A_1(T)$ in the undamped ($z=1$) and overdamped ($z=3$) cases. }
    \label{fig:intro}
\end{figure}

In this paper, we aim to resolve this issue by providing a new theory of quantum oscillations for a quantum critical metal at low temperatures beyond the LK paradigm. In principle, one could follow Luttinger's approach by carefully keeping track of the oscillatory parts of the fermion self-energy, which enters the thermodynamic potential.  Then, the remedy of the zero temperature divergence stated above would require an exact cancellation of the divergence in the thermodynamic potential, tracking which would be a rather involved task. Here we show that one can circumvent Luttinger's approach by utilizing the saddle point conditions for the thermodynamic potential. We present an exact formulation for  computing the entropy in a magnetic field of quantum critical metals within the Migdal-Eliashberg (ME) approximation and show that the entropy manifestly satisfies the laws of thermodynamics. As concrete examples, we considered two different types of models with parabolic dispersion and MFL behavior: one with overdamped bosons and a dynamical exponent $z=3$ and the other with undamped bosons and $z=1$. Although the $z=1$ model serves as a simple and elegant demonstration of the essentials of our calculation, the $z=3$ model is more physically relevant.  The comparison of the results based on these two models also unveils the important role of different dynamical scaling laws of the bosons in determining the oscillatory entropy and magnetization. Proceeding further, we obtain expressions for the entropy at temperatures and magnetic fields small compared to the bare Fermi energy of the metal, from which the expressions for the oscillation amplitudes $A_k(T)$ are obtained via the Maxwell relation. Our approach naturally captures the LK expressions in the non-interacting limit and results in new temperature dependence for the DHVA amplitude near a quantum critical point. Specifically, for $z=1$ bosons, we find that the low temperature behavior of $A_k(T)$ is $T^2 \ln T$, while for $z=3$ case, we find that the leading contribution to $A_k(T)$ is $T^{4/3}$. Both of these behaviors obey the thermodynamic laws, and deviate from the standard LK formula in Eq.\eqref{eq:LK1}. In addition, we find that the tail of $A_k(T)$ at temperatures greater than the cyclotron frequency $T\gg \omega_c$ is of the form $T\exp\{-\# T\ln T\}$ for both cases $z=1$ and $z=3$. These asymptotic limits are summarized in  Fig.~\ref{fig:intro}. Such apparent deviation from the LK formula could be, in principle, detected in experiments by measuring the quantum oscillations of either magnetization or specific heat, and can be used as a direct evidence for the existence of quantum critical point in various materials.     

With hindsight, our theory also provides a clear picture on where the recipe of Luttinger for the DHVA oscillations breaks down.  Perhaps not surprisingly, the temperature dependence of the fermion self-energy cannot be neglected near a quantum critical point, since it exhibits non-analytic behavior.  Additionally, we show explicitly  that the oscillatory parts of the self-energy cannot be neglected; due to their singular temperature dependence, they in fact can contribute even more strongly to the low-temperature entropy.  Lastly, we observe that the contribution from overdamped order parameter fluctuations in the metallic environment produces the dominant correction to the entropy, which in fact can contribute significantly to the temperature dependence of the DHVA amplitude. This is in sharp contrast to conventional metals (Fermi liquid regime) where the quantum oscillations are determined only by the fermions.

The outline of the paper is as follows. In Sec.~\ref{sec:ME} we summarize the ME theory of a marginal Fermi liquid in three dimensions and in the presence of a magnetic field. In Sec.~\ref{Sec:Luttinger_expansion} we recall the main steps in the derivation of the extended LK formula, and then demonstrate that, if applied to our theory at criticality, it leads to violations of the third law of thermodynamics. Next, in  Sec.~\ref{eq:LWE_potential} we consider the Luttinger-Ward-Eliashberg thermodynamic potential corresponding to our ME theory in a magnetic field. Starting from this representation, in Sec.~\ref{sec:Entropy_eq} we derive a formally exact formula for the entropy that manifestly satisfies the thermodynamic constraints.
This formula is then used to evaluate specific heat and magnetization oscillations in the particular case of the undamped critical boson in Sec.~\ref{sec:_z=1}, while the role of the Landau damping is discussed in Sec.~\ref{sec:_z=3}. Discussion and conclusions are presented in Sec.~\ref{sec:Discussion}. Some details of calculations are presented in the Appendices.


\section{Models of marginal Fermi liquid behavior}\label{sec:ME}
We will study DHVA oscillations in models exhibiting marginal Fermi liquid behavior in zero magnetic field, as it arises in a three dimensional metal tuned to a continuous quantum phase transition.  Broadly stated, there are two classes of symmetry breaking quantum phase transitions in metals in the absence of disorder: those that preserve lattice translations and those that break them.  Examples in the former category include ferromagnetic, Ising nematic transitions and more generally, Pomeranchuk instabilities.  Examples of the latter describe the onset of density wave order.  Within the former category, we can further distinguish between conserved order parameters (such as ferromagnetism) and non-conserved order parameters.  We will study transitions that preserve lattice translations associated with non-conserved order parameters.  

If such transitions are continuous, their universal properties are governed by a low energy theory consisting of fermions at finite density coupled to gapless  bosonic order parameter fluctuations by a Yukawa-like coupling, which is the most relevant coupling in the sense of the renormalization group.  A straightforward approach to such problems involves the use of Migdal-Eliashberg (ME) theory to obtain self-consistent descriptions  in terms of propagators of the fermions and bosons and their respective self-energies.  The bare fermion and boson propagators, $G_0 , D_0$ respectively, are
\begin{equation}\label{eq:bare_prop}
\begin{aligned}
\left[G_0(\ep_n,k) \right]^{-1}&=i\ep_n-\xi_k, \  \ \xi_k = \frac{k^2}{2m} - \mu, \\
\left[D_0(\omega_n,q) \right]^{-1}&=c^2q^2+\omega_n^2 + m_b^2.
\end{aligned}
\end{equation}
Here $\ep_n=2\pi T(n+1/2)$ and $\omega_n=2\pi T n$ are the fermionic and bosonic Matsubara frequencies respectively, $m$ is the fermionic mass, $c$ is the boson's velocity, and $\mu$ is the chemical potential. The boson mass $m_b$ is varied by a non-thermal parameter such as pressure, doping, etc., and the quantum critical point is accessed by tuning the physical boson mass to zero. We also note that Eq.~\eqref{eq:bare_prop} assumes a simple parabolic dispersion. We will comment on possible extensions in Sec.~\ref{sec:Discussion}.

The self-consistent relations involving these propagators are\cite{Eliashberg,Carbotte1990}
\begin{equation}
\label{selfconsistentzerofield}
\begin{aligned}
G(\ep_n,k)&= \frac{1}{[G_0(\ep_n,k)]^{-1}+i\Sigma(\ep_n,k)}  \\
D(\omega_n,q)&= \frac{1}{[D_0(\omega_n,q)]^{-1}-\Pi(\omega_n,q)}  \\
\Sigma(\ep_n,k)&= i g^2 T\sum\limits_m \int\frac{d^3q}{(2\pi)^3}D(\omega_m,q)G(\Omega_m+\ep_n,k+q) \\
\Pi(\omega_n,q)&=  -{g^2 T}\sum\limits_m \int\frac{d^3k}{(2\pi)^3}G(\omega_n+\ep_m,k+q)G(\ep_m,k).
\end{aligned}
\end{equation}  
The first two equations are the Dyson equations whereas the last two define the respective self-energies.  Originally used in the context of the electron-phonon problem, the ME theory neglects all quantum corrections of the boson-fermion vertex.  It arises as a saddle point of recent theories with random-flavor Yukawa couplings, studied in Ref. \cite{Esterlis2021,Haoyu2022}.  While this approximation can be justified by invoking formal large $N$ limits, it is sufficient also to assume, as is done in the electron-phonon context, that the characteristic boson speed is small compared to the Fermi velocity:  (i.e. if $\beta:= v_F/c \gg 1$ where $v_F=\sqrt{2\mu/m}$ is the Fermi velocity, or if the bosons are Landau overdamped), so that a version of the Migdal's theorem applies. For our purposes, we will take the ME equations above as the starting point for describing the quantum critical metal. \\

The self-consistent solution of Eqs.\eqref{selfconsistentzerofield} has been extensively studied before. Here we only briefly summarize its main properties, and the details of the calculation are presented in Appendix \ref{sec:self-energy}. First, the boson self-energy contains a Landau damping term\cite{PhysRevB.74.195126},
\begin{equation}
\Pi(\omega_n,q) \approx g^2\nu -\frac{\pi g^2\nu}{2v_F}\frac{|\omega_n|}{q}\;,\label{eq:Pi4}
\end{equation}
where $\nu=k_F^2/(2\pi^2v_F)$ is the density of states near the Fermi level. From this result, the $z=3$ dynamical scaling is readily recognized. Inserting this into the equation for $\Sigma(\varepsilon_n,k)$ and neglecting any weak momentum-dependence of the fermion self-energy near the Fermi surface, i.e. approximating $\Sigma(\varepsilon_n,k)\approx \Sigma(\varepsilon_n)$, the solution for $\Sigma(\varepsilon_n)$ can be found in a straightforward way.  In the low-temperature limit where $\varepsilon_n$ is treated as a continuous variable, $\Sigma(\varepsilon_n)$ contains a leading non-analytic temperature dependence $T\ln T$, with a subleading term of the order of $\mathcal{O}(T)$. Keeping only the leading term, we find, as $T\to0$,
\begin{equation}\label{eq:Sigma4}
     \Sigma(\ep_n)   \approx \frac{1}{3}\bar{g}^2\ep_n\ln\frac{\Lambda_D}{|\ep_n|}+\frac{1}{3}\bar{g}^2\ep_n +\frac{1}{3}\pi \bar{g}^2 T\ln \frac{T}{\Lambda_D},
\end{equation}
where $\bar{g}^2=g^2/(4\pi^2 c^2 v_F)$ is the dimensionless coupling constant, and $\Lambda_D= 2\beta \omega_D^3/(\pi g^2\nu)$ is the energy cutoff expressed in terms of the effective "Debye frequency", $\omega_D=E_F/\beta\ll E_F$. The first two terms of Eq.\eqref{eq:Sigma4} is the marginal Fermi liquid self-energy which was first discussed as a phenomenological model of the normal state of cuprate superconductors\cite{MFL}. Similar MFL self-energy can also be obtained in the undamped boson model, in which the boson self-energy is neglected and the dynamical exponent is, therefore, $z=1$. Following completely parallel calculations, we find that the fermion self-energy is identical to Eq.\eqref{eq:Sigma4} with $\bar{g}^2/3$ replaced with $\bar{g}^2$ and $\Lambda_D$ replaced with $\omega_D$ (see Appendix \ref{sec:self-energy} for calculation details). These expressions can also be analytically continued to the real frequency axis according to the following convention
\begin{equation}
    \begin{aligned}
        \Sigma^{R}(\omega)&=-i\Sigma(-i\omega),~~~ G^{R}(\omega,k)=\frac{1}{\omega-\xi(k)-\Sigma^R(\omega)}.
    \end{aligned}
\end{equation}
Here the superscript `R' denotes retarded functions which are analytic in the upper half-plane.

\subsection{Midgal-Eliashberg equations in a magnetic field}
To study DHVA oscillations in the marginal Fermi liquid described above, we will need to adapt the ME relations to allow for the presence of a magnetic field.  Since the  field reduces translation symmetry to a set of magnetic translations, the fermion propagators and self-energies (which are also gauge-dependent quantities) will no longer be translationally invariant.  This results in a modified Dyson equation, expressed in the position representation as
\begin{equation}\label{eq:Dyson_equation}
\begin{aligned}
      &i \int d^3\bm{r}''\;\Sigma(\bm{r}, \bm{r}'',\ep_m)G(\bm{r}'', \bm{r}',\ep_m)\\ &+\left[i\ep_m -\frac{\bm{\pi}^2}{2m}+\mu\right]G(\bm{r},\bm{r}',\ep_m)=\delta(\bm{r}-\bm{r}')\;,
    \end{aligned}
\end{equation}
where $\bm{\pi}=\bm{p}-e\bm{A}$ is the canonical momentum operator in the presence of a magnetic field $B \hat z$. Hereafter, we will adopt the Landau gauge in which the vector potential is $\bm{A}=(-yB,0,0)$.
By contrast, since the boson is taken to be electromagnetically neutral, both $D, \Pi$ remain translationally invariant.  
Consequently, the proper self-energy in the position representation  is expressed in terms of the self-consistency equation 
\begin{equation}\label{eq:SCBA_finite_B}
    \Sigma(\bm{r},\bm{r'},\ep_m) = ig^2 T\sum\limits_{m'}D(\ep_m-\ep_{m'},\bm{r}-\bm{r}')G(\bm{r},\bm{r}',\ep_{m'}).
\end{equation}
Using the properties of the bare fermionic Green's function in a magnetic field, as well as the structure of the self-consistency condition Eq.\eqref{eq:SCBA_finite_B}, it is easy to show that the self-energy can be factorized into a product of a phase factor and a translation-invariant part
\begin{equation}\label{eq:phase_SE}
\begin{aligned}
\Sigma(\bm{r},\bm{r}',\ep_m)&=e^{i\Phi(\bm{r},\bm{r}')} \bar{\Sigma}(\bm{r}-\bm{r}',\ep_m)\;,\\
\Phi(\bm{r},\bm{r}')&=(x'-x)(y+y')/2l_B^2\;,
\end{aligned}
\end{equation}
where $l_B$ is the magnetic length. This allows us to solve the Dyson equation \eqref{eq:Dyson_equation} exactly, without resorting to the explicit form of $\bar{\Sigma}(\bm{r}-\bm{r}',\ep_m)$. We find
\begin{equation}\label{eq:Gphase}
G(\bm{r},\bm{r}',\ep_m)=e^{i\Phi(\bm{r},\bm{r}')} \bar{G}(\bm{r}-\bm{r}',\ep_m),
\end{equation}
with the same phase factor as in Eq.~\eqref{eq:phase_SE}. 
The Fourier transform of the translation-invariant part of $G$ reads as
\be \label{eq:G_with_Sigma_k}
\bar{G}(\bm{k}, \ep_m)  =\sum\limits_{n=0}^{}2(-1)^n e^{-l_B^2k_\parallel^2}L_n\left(2l_B^2k_\parallel^2\right) G_n(k_z,\ep_m)\;,
\ee
where $n$ labels the Landau levels, $\omega_c$ is the cyclotron frequency, and $k_\parallel$ is the in-plane component of momentum perpendicular to the magnetic field. The Green's function for a given Landau level and $k_z$-momentum is denoted as $G_n(k_z,\ep_m)$, and given by
\begin{equation}\label{eq:G_for_n}
    G_n(k_z,\ep_m)=\frac{1}{i\ep_m-\omega_c(n+\frac{1}{2})-\frac{k_z^2}{2m}+\mu +i\Sigma_n(k_z,\ep_m)},
\end{equation}
and the associated self-energy $\Sigma_n(k_z,\ep_m)$ is related to the Fourier transform of $\bar{\Sigma}(\bm{r},\ep_m)$ defined in Eq.~\eqref{eq:phase_SE} as follows
\begin{equation}\label{eq:Sigma_n_vs_Bar_sigma}
   \Sigma_{n}(k_z,\ep_m)=\int\limits_0^{+\infty}dt e^{-t} L_n\left(2t\right)\bar{\Sigma}(\sqrt{t}/l_B,k_z,\ep_m).
\end{equation}
Here $L_n(t)$ is the Laguerre polynomial, and $\sqrt{t}/l_B$ replaces $k_\parallel$. Given the structure of the full Green's function Eqs.~(\ref{eq:Gphase},~\ref{eq:G_with_Sigma_k}), the bosonic self-energy can be then conveniently written as 
\begin{equation}\label{eq:Pi_eq}
\begin{aligned}
   &\Pi(\omega_{\bar{m}},q)=  -g^2 T\sum\limits_m \hspace{-0.3em} \int\hspace{-0.3em}\frac{d^3k}{(2\pi)^3}\bar{G}(\omega_{\bar{m}}+\ep_m,k+q)\bar{G}(\ep_m,k)\\
   &=-g^2 T\sum\limits_{\substack{m\\n\bar{n}}}  \frac{X_{n\bar{n}} (q_\parallel)}{(2\pi l_B)^2} \hspace{-0.2em}\int\limits_{k_z} \hspace{-0.2em} G_n(k_z+q_z,\ep_m+\omega_{\bar{m}})G_{\bar{n}}(k_z,\ep_m)
   \end{aligned}
\end{equation}
Also, $\int_{k_z}$ in the second line in Eq.~\eqref{eq:Pi_eq} stands for the integral over $k_z$, and we defined the matrix element $ X_{n\bar{n}}(q)$ as
\begin{equation}\label{eq:A_nnn}
  X_{n\bar{n}}(q)= \left(-1\right)^{\bar{n}-n} e^{-\frac{l_B^2q^2}{2}} L_n^{\bar{n}-n}\left(\frac{l_B^2q^2}{2}\right) L_{\bar{n}}^{n-\bar{n}}\left(\frac{l_B^2q^2}{2}\right).
\end{equation}
We also note that for high Landau levels and for the typical momenta such that $q\ll k_F$, this form-factor can be approximated as
\begin{equation}\label{eq:B_nnn}
     X_{n\bar{n}}(q)\approx J_{|n-\bar{n}|}^2(R_c q)\;,\quad  q\ll k_F\;,
\end{equation}
where $J_n(x)$ is the Bessel function, and $R_c=v_F/\omega_c$ is the cyclotron radius. Finally, the bosonic self-energy is related to the full bosonic propagator in the same way as in the second line of Eq.~\eqref{selfconsistentzerofield}. It is convenient to use Eq.~\eqref{eq:Sigma_n_vs_Bar_sigma} and transform the self-consistency condition  \eqref{eq:SCBA_finite_B} into a closed-form equation that contains $\Sigma_n$ only
\begin{equation} \label{eq:Sigma_eq_new}
\begin{aligned}
\Sigma_n(\ep_m,k_z)&=ig^2 T\sum\limits_{\bar{m},n} \int\frac{d^3q}{(2\pi)^3} D(\ep_m-\ep_{\bar{m}},q) \times\\
&\times X_{n\bar{n}}(q_\parallel)  G_{\bar{n}}(k_z+q_z,\ep_{\bar{m}})\;.
\end{aligned}
\end{equation}
Eqs.~(\ref{eq:Pi_eq},~\ref{eq:Sigma_eq_new}) in combination with  Eq.~(\ref{eq:G_for_n}) and the second line of Eq.~\eqref{selfconsistentzerofield} form a closed system generalizing the standard one-loop  equations \eqref{selfconsistentzerofield} to the finite magnetic field case. We will make use of these relations in Sec.~\ref{sec:Entropy_eq} when we evaluate the entropy in a magnetic field of the marginal Ferm liquid.  We emphasize that these equations are valid up to all orders in both $\omega_c/\mu$ and the dimensionless coupling strength, as long as the Migdal approximation still holds. On the technical level, this means that the vertex corrections are negligible. Similarly to the conventional electron-phonon problem (see \cite{CHUBUKOV2020168190} for a recent critical analysis of the ME approximation there), this requires temperatures much lower than some characteristic (`Debye') temperature scale at which the critical order parameter becomes strongly Landau-overdamped due to its fast decay into a particle-hole continuum.
This scale should also be much smaller than the Fermi energy $\mu$ so that the bosonic degrees of freedom are effectively much slower than the fermionic ones. In addition, the presence of the magnetic field requires that the cyclotron frequency is much smaller than the Debye frequency so that the boson is still overdamped on the scale of the magnetic length. Therefore, we expect the ME approach to remain valid in the present problem provided two conditions: (i) $T,\omega_c\ll \omega_D$, and (ii) $\omega_D \ll \mu$, with not too strong interactions.
 

\section{Extended LK framework and its breakdown}\label{Sec:Luttinger_expansion}
Before constructing a theory of DHVA oscillations of marginal Fermi liquids described in the previous section, we recount the standard approach to DHVA oscillations and show how they break down at a quantum critical point.  Since the breakdown of the extended LK approach to DHVA is not specific to the ME approximation, we present a more general discussion in what follows. 

A clean, three-dimensional spherically symmetric Fermi gas in a magnetic field $B \hat z$  has the following thermodynamic potential per unit volume ($\hbar =  k_B = 1$):
\begin{equation}
\Xi^0 = - \frac{m \omega_c}{2 \pi} T \int \frac{d p_z}{2 \pi} \sum_n\log{\left[1 + e^{ \left( \mu-\epsilon_{np_z} \right)/T} \right]},
\end{equation}
where $\epsilon_{np_z} = \frac{p_z^2}{2m} + \omega_c (n+ 1/2)$.  
Upon summing over the Landau levels via the Poisson formula \footnote{We refer to the first term in this summation formula as the `smooth' part, whereas the second term is denoted as the `oscillatory' part.},
\begin{equation}\label{eq:Poisson}
    \begin{aligned}
\sum\limits_{n=0}^{+\infty}F_{n+\frac{1}{2}}=\int\limits_0^{+\infty} dx  F_x
+2\sum\limits\limits_{k=1}^{+\infty}(-1)^k\hspace{-0.2em}  \int\limits_0^{+\infty} dx \cos\left(2\pi k x\right) F_x,
\end{aligned}
\end{equation}
and performing the $p_z$ integration via saddle point methods, one finds the following expression for the oscillatory component of the thermodynamic potential \cite{lifshitz1956theory,abrikosov2017fundamentals}:
\begin{equation}\label{eq:Xi_0_osc_form}
\Xi^0_{osc}(T) = \frac{\left(m \omega_c \right)^{3/2}}{2 \pi^2} \sum_{k=1}^{\infty} \frac{\left( -1 \right)^k}{k^{3/2}} A_k(T) \cos{\left( \frac{2 \pi \mu k}{\omega_c} - \frac{\pi}{4}  \right)},
\end{equation}
where $A_k$, the amplitude of the k$^{th}$ harmonic of the non-interacting system is given by Eq.~\eqref{eq:LK1}. Throughout the rest of the paper, we are going to use the terms `smooth', `non-oscillatory', or `uniform' interchangeably to denote contributions to various quantities (i.e. observables, self-energies, etc.) that are smooth functions of the magnetic field. Such contributions are typically produced by the first term in the Poisson summation formula, Eq.\eqref{eq:Poisson}. On the other hand, the remaining contributions that exhibit oscillations as a function of $\mu/\omega_c$ will be referred to as `oscillatory' terms which originate from the second term on the r.h.s. of Eq.\eqref{eq:Poisson}.
Upon introducing the scaling variable $\lambda = 2 \pi^2 T/\omega_c$, it becomes evident that  $\Xi_{osc}/\Xi_{n-osc} \sim (\omega_c/\mu)^{5/2}$, where $\Xi_{n-osc}$ denotes the non-oscillatory piece of the thermodynamic potential.  Next, including interactions, the thermodynamic potential for a system with Hamiltonian $H=H_0 + H_{int}$  (and $H_0$ is the free Fermi gas Hamiltonian), namely
\begin{equation}
\Xi = \Xi^0 - T \log{\langle T_{\tau} e^{-\left[ \int_0^T d \tau H_{int}(\tau) \right]} \rangle }, 
\end{equation}
can equivalently be expressed as \cite{LuttingerWard}
\begin{equation}
\label{luttingertp}
\Xi = -T \sum_{m}  {\rm Tr} \left[  \log{\left[-\hat G(\varepsilon_m)^{-1} \right]}  - i \hat G(\varepsilon_m)  \hat \Sigma(\varepsilon_m) \right] + \Phi,
\end{equation}
where the Green function $\hat G$ and  electron self-energy $\hat \Sigma$ are matrix-valued, the trace operation takes care of momentum integration and Landau level summation, and the regularization factor $e^{i \varepsilon_m 0^+}$ is assumed. The last term $\Phi$ is the so-called Luttinger-Ward functional, which involves an infinite set of closed connected skeleton diagrams (i.e. those that have no self-energy insertions but with fully dressed propagators on each internal line).  It can be shown that the expression for $\Xi$ in Eq.~\eqref{luttingertp} is stationary with respect to variation of the exact self-energy.  

In a magnetic field, the self-energy will consist of a uniform piece and an oscillatory piece: $\hat \Sigma(\varepsilon_m) = \hat \Sigma^{(n-osc)}(\varepsilon_m) + \hat \Sigma^{(osc)}(\varepsilon_m)$.  In three dimensions, $\hat \Sigma^{(osc)}$ is smaller than its non-oscillatory component by a factor $(\omega_c/\mu)^{3/2}$, provided that the interactions responsible for self-energy corrections are sufficiently short-ranged (as an example, in Appendix \ref{sec:phonons} we explicitly demonstrate this suppression in case of the electron-phonon interaction). Thus, expanding the thermodynamic potential in powers of $\hat \Sigma^{(osc)}$ and making use of the stationary condition, the correction to the thermodynamic potential from the oscillatory self-energy is $\mathcal O( [ \hat\Sigma^{(osc)} ]^2 ) \sim (\omega_c/\mu )^3$.  It is thus parametrically smaller than the bare LK expression above, and can be neglected.  

Upon neglecting all contributions from $\hat \Sigma^{(osc)}$, it is a simple matter to convince oneself that a cancellation occurs between the second term in Eq.~\eqref{luttingertp} and $\Phi$.  We therefore arrive at Luttinger's starting expression from which to extract DHVA oscillations:
\begin{equation}
\Xi_{osc} \approx -T \sum_{m} e^{i \varepsilon_m 0^+} {\rm Tr} \left[  \log{\left[-\hat G_0(\varepsilon_m) - i\hat\Sigma^{(n-osc)}(\varepsilon_m) \right]} \right].
\end{equation}
After performing the Landau level summation and the $k_z$-integral, we can extract the expression for the oscillation amplitudes,
\begin{equation}\label{eq:A_K}
    A_k= T\sum\limits_{m\geq 0} \exp\left\{-\frac{2\pi k }{\omega_c} \Big(\ep_m+\Sigma(\ep_m,k_F,k_z=0)\Big) \right\}.
\end{equation}
Here we assumed that $\hat\Sigma^{(n-osc)}(\varepsilon_m)$ is only weakly dependent on the magnetic field, and thus, can be diagonalized in the momentum basis as $\Sigma(\ep_m,k_\parallel,k_z)$ at the leading order in $\omega_c/\mu$ \cite{GorkovBychkov}. Note that by setting $\Sigma=0$ we immediately arrive at Eq.\eqref{eq:LK1}.

For later convenience, we convert the Matsubara sum above into a real frequency integral \footnote{We use the following sign convention for analytic continuation: $i \epsilon_m \rightarrow \omega + i \delta, i \Sigma(\epsilon_m) \rightarrow -i \Sigma(-i \omega) = \Sigma^{Ret}(\omega)$.  }, to obtain the following expression for the amplitude:
\begin{eqnarray}
\label{ampextendedlk}
A_k = &&\int_{-\infty}^{\infty} \frac{d \omega}{2 \pi} e^{\omega 0^+}n_F(\omega) \exp{\left[\frac{2 \pi k}{\omega_c} {\rm Im} \Sigma^{R}(\omega) \right] } \nonumber \\
&&\times \sin{\left(\frac{2 \pi k}{\omega_c} \left[ \omega - {\rm Re} \Sigma^{R}(\omega) \right] \right)},
\end{eqnarray}
where the self-energy appearing above is the non-oscillatory self-energy.  
In the absence of self-energy corrections, the expression above reproduces the LK form for the amplitude.  We next discuss thermodynamic constraints on the expression above in the case of interacting systems.

\subsection{Thermodynamic constraints}
The third law of thermodynamics requires that the entropy $S = - \partial \Xi/\partial T$ and the specific heat $C=-T \partial^2 \Xi/\partial T^2$ must both vanish in the zero temperature limit.  For this to hold in the presence of the field, the DHVA amplitudes must satisfy 
\begin{equation}\label{eq:constraints}
\lim_{T \rightarrow 0} \frac{\partial A_k}{\partial T} = 0, \ \lim_{T \rightarrow 0} T \frac{\partial^2 A_k}{\partial T^2} = 0.
\end{equation}
These constraints are trivially obeyed for case of the free Fermi gas [Eq.~\eqref{eq:LK1}].  With interactions treated within Luttinger's approach, we must consider derivatives with respect to $T$ of the amplitude given in Eq.~\eqref{ampextendedlk}.  Temperature dependence in this expression arises from $n_F(\omega)$ and $\Sigma^{R}(\omega)$.   In the first case, temperature derivatives of $n_F(\omega)$ are sharply peaked at $\omega=0$ in a width of the order of $T$, which guarantees that the integral vanishes in the low temperature limit.  The second contribution contains temperature derivatives of $\Sigma^{R}(\omega)$.  In a generic Fermi liquid, such derivatives vanish at $T=0$.  For instance, the imaginary part of the self-energy in a Fermi liquid is typically of the form ${\rm Im }\Sigma^{R}(\omega) \sim T^2 + \omega^2$, and thus, $ \left( \partial {\rm Im} \Sigma^{R}(\omega)/\partial T \right)_{T=0} = 0$.  

Now, let us imagine using these approximations in the case of a marginal Fermi liquid, described in the previous section.  In addition to the expression above, we would have a contribution to the thermodynamic potential from the bosonic sector.  But let us proceed in a cavalier fashion and ignore this contribution, since afterall, the bosonic fluctuations are neutral.  Later on, we will show that surprisingly, the bosonic contribution is the dominant one, and cannot be neglected, but for now, we proceed naively and ignore the bosons.  
As we showed in Section II, the self-energy typically of a marginal Fermi liquid exhibits non-analytic temperature dependence:
\begin{equation}
{\rm Im} \Sigma^{R}(\omega) \approx - \bar g^2  \frac{\pi }{2} \vert \omega \vert - \pi \bar g^2 T \log{\left( \frac{\beta T}{\vert \omega \vert} \right)},\label{eq:self28}
\end{equation}
where $\bar g, \beta$ are constants defined in the previous section.  Clearly, in the zero temperature limit,
\begin{equation}
\lim_{T \rightarrow 0} \frac{\partial {\rm Im}\Sigma^{R}(\omega)}{\partial T}   \approx -\pi \bar g^2  - \pi \bar g^2 \log{\left( \frac{\beta T}{\vert \omega \vert} \right)}
\end{equation}
results in the divergence in $\partial A_k(T)/\partial T$ as $T \rightarrow 0$.  More specifically, as shown in the appendix, the asymptotic evaluation of the expression in Eq.~\eqref{ampextendedlk} leads to the following low temperature dependence: 
 \begin{equation}\label{eq:A_kT=0}
A_k(T=0) \approx \frac{\omega_c}{4\pi^2 \bar{g}^2  k \ln (2\pi k \frac{\omega_D}{\omega_c })} \left[1+\mathcal{O}\left(\frac{\ln \ln (k \omega_D/\omega_c)}{\ln (k \omega_D/\omega_c)}\right)\right],
\end{equation}
and
\begin{equation}\label{eq:Extended_LK_A}
 \begin{aligned}
     A_k &\approx A_k(T=0)+ A_k^{(a)}+ A_k^{(b)}, \\	
    A_k^{(a)}&\approx -\frac{\pi^2\bar{g}^2 k T^2}{6\omega_c}\ln \frac{\omega_D}{2\pi T}+\mathcal{O}(T^2),\\
    A_k^{(b)}&\approx   \frac{2\pi^2 \bar{g}^2 k T}{\omega_c}\ln \frac{\omega_D}{\beta T}\; A_{k}(T=0) +\mathcal{O}(T).
        \end{aligned}
\end{equation}
These non-analyticies could be regularized by keeping the correlation length finite (i.e. by introducing a finite boson mass $m_b>0$). In this case, the thermal part of the self-energy crosses over to $\sim T^2$ at sufficiently low temperatures (i.e. outside of a quantum critical fan), where we expect the extended LK scheme again to work  well.

\begin{figure}
    \centering
    \includegraphics[width=0.95\columnwidth]{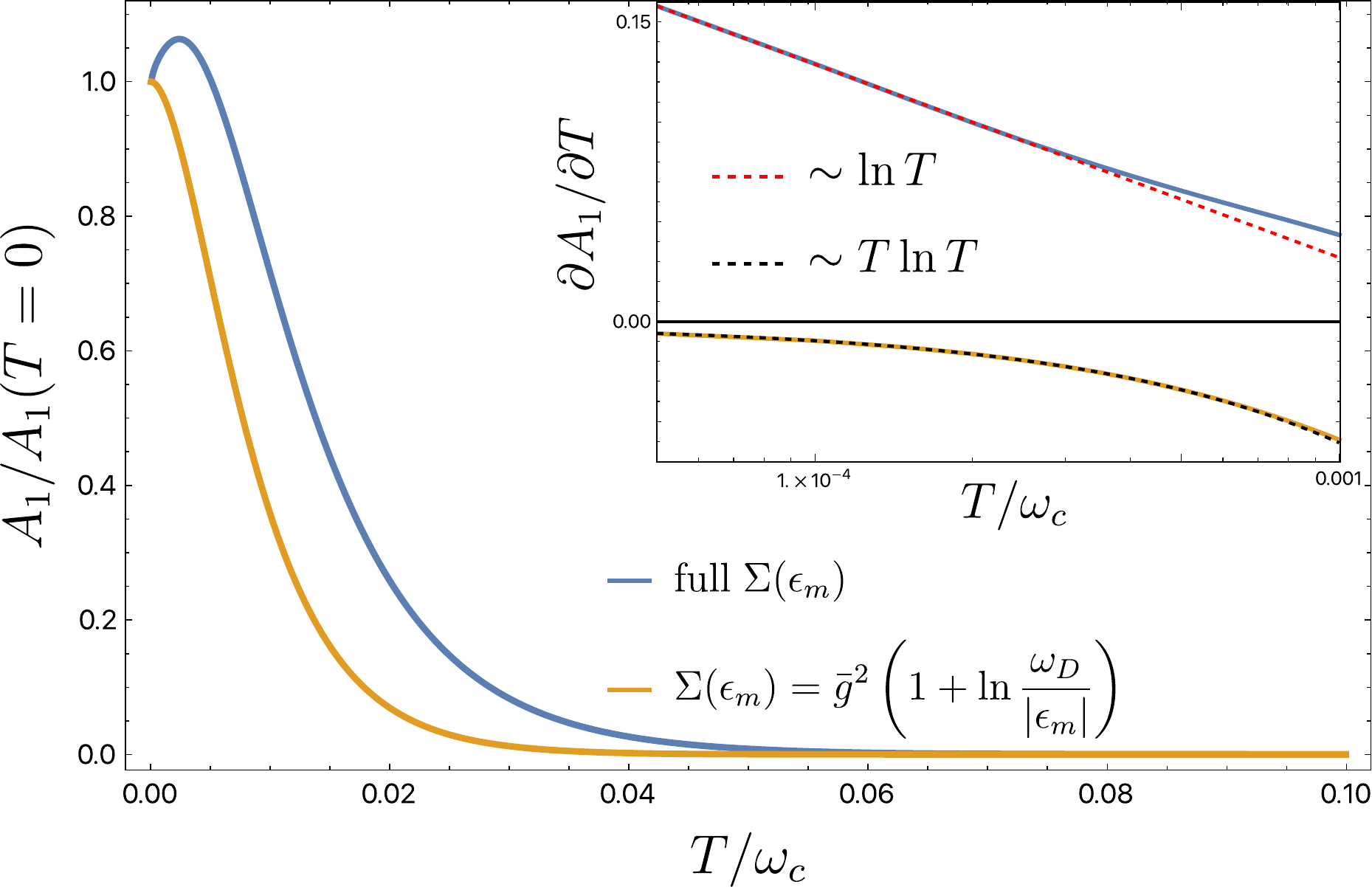}
    \caption{Numerical evaluation of the first oscillation amplitude $A_1(T)$ and its temperature derivative $\partial A_1/\partial T$ (inset), based on the extended LK formula Eq.\eqref{ampextendedlk}. For comparison, two different expressions for the fermionic self-energy are used: the full fermion self-energy Eq.~\eqref{eq:Sigma_z=3_B=0_full}, and the marginal Fermi liquid self-energy in the absence of any thermal effects $\Sigma(\ep_m)=\bar{g}^2\ep_m\left(1+\ln \omega_D/|\ep_m|\right)$. We note that $A_1(T=0)$ is the same for both cases. The parameters used are $\bar{g}^2=6$, $\Lambda_D=10\omega_c$, $v_F/c=100$.}
    \label{fig:AT}
\end{figure}

In Fig.~\ref{fig:AT} we show the numerically evaluated first harmonic amplitude $A_1(T)$ and its temperature derivative $\partial A_1/\partial T$ based on the extended LK formula Eq.~\eqref{ampextendedlk}, and the self energy is calculated from Eq.~\eqref{eq:Sigma_z=3_B=0_full}. Here we choose $\beta\equiv v_F/c=100$, $\omega_D=10\omega_c$, and set $\bar{g}^2=6$. All the energy scales are measured in units of the cyclotron frequency. For comparison, we show two different results based on two different self-energies. The blue curve is for the result obtained by using the full MFL self-energy, whereas the orange curve is for the result obtained by explicitly excluding the thermal contributions [the last term in Eq.\eqref{eq:self28}] from the self-energy. 
From the inset in Fig.~\ref{fig:AT}, one can clearly see that the temperature derivative of $A_1(T)$ diverges logarithmically at small $T$ when the full self-energy is used. This leads us to conclude that the divergence is caused by the explicit non-analytic $\sim T\ln T$ temperature dependence of the self-energy. Thus, it is manifest that the naïve application of Eq.~\eqref{ampextendedlk} with our NFL self-energy leads to a divergence of $\partial A_1/\partial T$ as $T\to0$. This is consistent with our analytical results in Eq.~\eqref{eq:Extended_LK_A} above.

Although the extended LK scheme breaks down for the marginal FL at low temperatures, it works perfectly well in the opposite regime $T \gg \omega_c$ (but still $T\ll \omega_D$), where finite temperature smearing of oscillations occurs. In this case, one can easily see that Eq.~\eqref{eq:A_K} is dominated only by the first Matsubara frequency $\ep_0=\pi T$ term. Thus, we recover the standard result \cite{FowlerPrange}
\begin{equation}
    A_k\approx T\exp\left\{-\frac{2\pi k }{\omega_c} \Big(\pi T+\Sigma(\pi T)\Big) \right\}, \quad T\gg \omega_c.
\end{equation}
In our marginal Fermi liquid model, we thus obtain (see Eq.~\eqref{eq:Sigma_first_Matsubara} and Appendix \ref{sec:self-energy} for details)
\begin{equation}\label{eq:A_high_T}
   A_k\approx T\exp\left\{-\frac{2\pi^2 \bar{g}^2 k T}{\omega_c}  \ln \left(\frac{\beta \omega_D}{\pi\bar{g}^2 T}\right)\right\},
\end{equation}
where we neglected the bare $\pi T$ contribution compared to $\Sigma(\pi T)$. Therefore, we see that the oscillations are strongly damped by thermal smearing at high temperatures $T\gg \omega_c$, and are qualitatively different than the smearing in a Fermi liquid, given by Eq.~\ref{eq:LK1}. We emphasize that $\Sigma(\pi T)$ is contributed only by the static mode - the `dynamical' part of the self-energy (i.e. the inelastic scattering rate) always vanishes at the first Matsubara frequency \cite{FowlerPrange,Maslov2003,Chubukov_first_Matsubara}. 

Thus, to obtain the physically correct low-temperature ($T\ll \omega_c$) form of DHVA oscillations at a quantum critical point, one must proceed with a more careful analysis of the full thermodynamic potential, including the contribution from  the oscillatory part of the self-energy, and from $\Phi$, the Luttinger-Ward functional.  Also, as we demonstrate below, non-trivial dynamical scaling laws also contribute and cannot be ignored.  In the Migdal-Eliashberg limit, all these effects can be incorporated.  


\section{Luttinger-Ward-Eliashberg thermodynamic potential}\label{eq:LWE_potential}
In this section, we return to the marginal Fermi liquid in the ME approximation.  For a generic coupled electron-boson system, the corresponding thermodynamic potential, viewed as a variational free energy functional of propagators and self-energies, yields the self-consistent relations as a saddle point condition \cite{Eliashberg}. A key advantage of the ME approximation is that the Luttinger-Ward functional $\Phi$ becomes especially simple and could be expressed as a single skeleton diagram (rather than an infinite number of them), such that the resulting saddle point conditions coincide with the ME equations. A thermodynamic potential having these properties can be written as the sum of three different contributions: 
\begin{equation}
\Xi = \Xi_{\text{el}} + \Xi_{\text{bos}} + \Phi,
\end{equation}
where the fermionic part $\Xi_{\rm el}$ is given by
\begin{equation}
\begin{aligned}
    &\Xi_{\rm el} =-T\sum\limits_{m}\ln\operatorname{det} \left(( \bm{\pi}^2/2m-\mu -i\ep_m )\delta(
   \bm{r}-\bm{r}') \right.\\ &- i\Sigma(\bm{r},\bm{r}',\ep_m)) 
    +iT\sum\limits_{m}\int d\bm{r}d\bm{r}' G(\bm{r},\bm{r}',\ep_m)\Sigma(\bm{r}',\bm{r},\ep_m)
    \end{aligned}
\end{equation}
Here the fermionic Green's function and the self-energy  are expressed in the coordinate-frequency representation. The bare Green's function $G_0$ is defined through the standard equation $(i\ep_m -\bm{\pi}^2/2m+\mu)\delta(
    \bm{r}-\bm{r}') = G_0^{-1}(\bm{r},\bm{r}',\ep_m)$.

    The bosonic contribution to the thermodynamic potential $\Xi_{\rm bos}$ has the following form
\begin{equation}
\begin{aligned}\label{eq:Xi_boson2}
    \Xi_{\rm bos}&=\frac{1}{2}T\sum\limits_{m}\int\frac{d^3q}{(2\pi)^3}\ln\left([D_0(\omega_m,q)]^{-1}-\Pi(\omega_m,q)\right)\\
    &+\frac{1}{2}T\sum\limits_{m}\int\frac{d^3q}{(2\pi)^3}\Pi(\omega_m,q)D(\omega_m,q),
    \end{aligned}
\end{equation}
The bare propagator $D_0$ is defined in the same way as in Eq.~\eqref{eq:bare_prop}. Finally, the Luttinger-Ward functional $\Phi$ in the ME approximation captures the interaction between the bosons and
fermions, and is given by 
\begin{equation}\label{eq:Xi_int}
\begin{aligned}
    \Phi&=\frac{g^2 T^2}{2}\sum\limits_{mm'}\int d\bm{r}d\bm{r}'D(\ep_m-\ep_{m'},\bm{r}-\bm{r}')\times \\
&\times G(\bm{r},\bm{r}',\ep_m)G(\bm{r}',\bm{r},\ep_{m'})\;.
    \end{aligned}
\end{equation}
 At zero magnetic field, the thermodynamic potential can be evaluated in the momentum basis, yielding the following expression for the fermionic part 

\begin{equation}
\begin{aligned}
\Xi_{el} &=-T\sum\limits_{n}e^{i\ep_n 0^+} \hspace{-0.5em}\int\frac{d^3k}{(2\pi)^3}\ln \left(-[G_0(\ep_n,k)]^{-1}-i\Sigma(\ep_n,k)\right)\\ & +iT\sum\limits_{n} e^{i\ep_n 0^+}\hspace{-0.4em}\int\frac{d^3k}{(2\pi)^3} G(\ep_n,k)\Sigma(\ep_n,k),
\end{aligned}
 \end{equation}
and the Luttinger-Ward functional reads as 
\begin{equation}
 \Phi  = \frac{g^2 T^2}{2}\sum\limits_{mn}\int\frac{d^3kd^3q}{(2\pi)^6} D(\omega_m,q)G(\Omega_m+\ep_n,k+q)G(\ep_n,k).
\end{equation}
Viewing $\Xi$ above as a functional of $G,D, \Pi, \Sigma$, it is evident that the first two relations in Eq. \ref{selfconsistentzerofield} follow from $\partial \Xi/\delta \Pi = 0$ and $\delta \Xi/ \delta \Sigma = 0$, whereas the latter two relations in Eq. \ref{selfconsistentzerofield} derive from $\partial \Xi/\delta G = 0$ and $\delta \Xi/ \delta D = 0$.

In a similar way, 
the thermodynamic potential $\Xi$ in a finite magnetic field results in Eqs.~(\ref{eq:Dyson_equation},\ref{eq:SCBA_finite_B},\ref{eq:Pi_eq})
upon varying it with respect to propagators and self-energies. One could also directly express the thermodynamic potential in the basis of Landau levels.  The fermionic part is then given by
\begin{equation}
\begin{aligned}
    \Xi_{\rm el} &=\frac{m\omega_c}{2\pi}T\sum\limits_{m}\sum\limits_{n=0}^{+\infty}\int\limits_{-\infty}^{+\infty}\frac{dk_z}{2\pi}\Big\{ \ln\left(-G_n\left(\ep_m,k_z\right)\right)  \\
    &+i\Sigma_n \left(\ep_m,k_z\right)G_n\left(\ep_m,k_z\right)\Big\},
    \end{aligned}
\end{equation}
and the electron-boson interaction (the Luttiner-Ward functional) takes the form 
\begin{equation}
\begin{aligned}\label{eq:Phi_mag_field}
    \Phi &=\frac{g^2 m\omega_c }{4\pi}T^2\sum\limits_{m\bar{m}}\sum\limits_{n\bar{n}=0}^{+\infty} \int\frac{d^3q}{(2\pi)^3}D(q,\ep_{\bar{m}}) X_{n\bar{n}}(q_\parallel)\\
    &\times \int\limits_{-\infty}^{+\infty}\frac{dk_z}{2\pi} G_n(\ep_m+\omega_{\bar{m}},k_z+q_z) G_{\bar{n}}(\ep_m,k_z),
    \end{aligned}
\end{equation}
with the Green's functions $G_{n}(\ep_m,k_z)$ and self-energies $\Sigma_n \left(\ep_m,k_z\right)$ defined for a given Landau level, and related to each other via Eq.~\eqref{eq:G_for_n}. Variations of $\Xi$ with respect to these propagators directly lead to the Eq.~\eqref{eq:Sigma_eq_new} and the second line of Eq.~\eqref{eq:Pi_eq}.

At this point, there are multiple ways of how one could proceed. For instance, by following the logic outlined in \cite{ChubukovFreeEnergyQCP}, one could note that the second term in the bosonic part of the thermodynamic potential Eq.~\eqref{eq:Xi_boson2} exactly cancels the LW (interaction) term $\Phi$, Eq.~\eqref{eq:Phi_mag_field}. As a result, $\Xi$ can be written in the following form 
\begin{equation}
\begin{aligned}
    \Xi &=\frac{1}{2}T\sum\limits_{m}\int\frac{d^3q}{(2\pi)^3}\ln\left([D_0(\omega_m,q)]^{-1}-\Pi(\omega_m,q)\right)\\
    &+
    \frac{m\omega_c}{2\pi}T\sum\limits_{m}\sum\limits_{n=0}^{+\infty}\int\limits_{-\infty}^{+\infty}\frac{dk_z}{2\pi}\Big\{ \ln\left(-G_n\left(\ep_m,k_z\right)\right)  \\
    &+i\Sigma_n \left(\ep_m,k_z\right)G_n\left(\ep_m,k_z\right)\Big\}.
    \end{aligned}
    \end{equation}
Next, if $\Sigma_n(\ep_n,k_z)$ does not have any essential dependence on the Landau level $n$ then the `smooth' contributions to the Poisson summation in the last two terms cancel each other, and we obtain
\begin{equation}\label{eq:Xi_An}
\begin{aligned}
\Xi&=\frac{(m\omega_c)^{3/2}}{2\pi^2} \sum\limits_{k=1}^{+\infty} \frac{(-1)^k}{k^{3/2}}  B_k \cos\left(\frac{2\pi k \mu}{\omega_c}  -\frac{\pi}{4}\right)\\
&+\frac{1}{2}T\sum\limits_{m}\int\frac{d^3q}{(2\pi)^3}\ln\left([D_0(\omega_m,q)]^{-1}-\Pi(\omega_m,q)\right).
\end{aligned}
\end{equation}
Note that the bosonic term also contains oscillatory contributions via oscillations of $\Pi(\omega_m,q)$. 
The amplitude in the first term is given by
\begin{equation}\label{eq:Xi_Bn}
    B_k= T\sum\limits_{m\geq 0}\left(1+\frac{2\pi k}{\omega_c} \Sigma(\ep_m,0)\right) e^{-\frac{2\pi k }{\omega_c} \Big(\ep_m+\Sigma(\ep_m,0)\Big)}.
\end{equation}
However, Eq.~\eqref{eq:Xi_An} and Eq.~\eqref{eq:Xi_Bn} are not very convenient for extracting the low-temperature behavior of the oscillation amplitudes since here one has to evaluate the self-energy dependence on temperature $T$. Instead, one can make the temperature dependence more manifest by considering the low-$T$ asymptotic of the entropy instead of the thermodynamic potential itself, which we derive in the next section. The DHVA magnetization can be then obtained via a Maxwell relation. To this end, we analytically continue the expression for the thermodynamic potential and express it in terms of real frequencies. Doing so, we find
\begin{widetext}
\begin{equation} \label{eq:thermodynamicpotential_general_1}
\begin{aligned}
    \Xi &= \frac{m\omega_c}{2\pi^2}\int\limits_{-\infty}^{+\infty}d\Omega n_F(\Omega)\sum\limits_{n=0}^{+\infty}  \int\limits_{-\infty}^{+\infty}\frac{dk_z}{2\pi}\left\{\operatorname{Im}\ln\Big(-[G^R_n\left(\Omega,k_z\right)]^{-1}\Big) + \Im \left[ G^R_n(\Omega,k_z)\Sigma^R_n(\Omega,k_z) \right] \right\}
   \\
  &+\int\frac{d\Omega}{2\pi}  n_B(\Omega) \int\frac{d^3q}{(2\pi)^3}\Big\{\operatorname{Im}\ln\left([D^R(\Omega,q)]^{-1}\right)+\operatorname{Im} \left[ \Pi^R(\Omega,q) D^R(\Omega,q) \right]\Big\}  + \Phi,\\
   \end{aligned}
\end{equation}
 where $\Phi$ can be expressed in terms of the fermion and boson spectral functions as
\begin{equation}
\begin{aligned}\Phi&=\frac{m\omega_c g^2}{8\pi^6}\sum\limits_{n\bar{n}}\int\frac{d^3q}{(2\pi)^3}X_{n\bar{n}}(q_\parallel)\int\limits_{-\infty}^{+\infty}\frac{dxdyd\Omega d k_z}{x-\Omega-y} \operatorname{Im}D^R(y,q)\operatorname{Im}G_n^R(x,k_z)\operatorname{Im}G_{\bar{n}}^R(\Omega, k_z+q_z)
\\&\times\Big(n_F(x)-n_F(\Omega+y)\Big)\Big(1+n_B(y)-n_F(\Omega)\Big).
  \end{aligned}
\end{equation}
\end{widetext}

Thus, starting with a thermodynamic potential whose saddle point yields the Migdal-Eliashberg self-consistency relations, we obtained an alternate expression for the thermodynamic potential expressed in the basis of Landau levels, which will prove to be more convenient in analyzing DHVA oscillations.  In the next section, we obtain an exact expression for the entropy, from which the low-temperature behavior of the DHVA oscillations can be extracted via a Maxwell relation.  


\section{Exact expression for the Migdal-Eliashberg entropy in a magnetic field}\label{sec:Entropy_eq}
In this section, we derive the exact expression for the entropy that only involves temperature derivatives of the distribution functions. Our derivation extends the result of \cite{vanderheyden1998self} to the finite magnetic field case. Starting with the expression for the thermodynamic potential in terms of real frequencies derived in the previous section, we make use of the following schematic relation
    \begin{equation}\label{eq:S_chain}
    S=-\frac{\partial\Xi}{\partial T}= -\left(\frac{\partial' \Xi}{\partial' T} \right)_{\Sigma,\Pi }-\underbrace{\frac{\delta \Xi}{ \delta \Sigma } }_{=0}\frac{\partial \Sigma}{\partial T}-\underbrace{\frac{\delta \Xi}{ \delta \Pi } }_{=0}\frac{\partial \Pi}{\partial T}\;,
    \end{equation}
    where $\Xi$ is viewed as a functional of $\Sigma$ and $\Pi$ (assuming that $G$ and $D$ are expressed in terms of $\Sigma$ and $\Pi$ via the Dyson equation). The partial derivative $\partial'/\partial'T$ in the first term acts only on the distribution functions while keeping $\Sigma$ and $\Pi$ fixed. The derivatives of $\Sigma$ and $\Pi$ do not contribute due to the stationary condition on $\Xi$ \cite{vanderheyden1998self}.

    Next, one can easily check that $\partial' \Phi/\partial'T$ cancels the corresponding terms in Eq.~\eqref{eq:thermodynamicpotential_general_1} involving the combinations $\operatorname{Im}[G_n^R(\Omega,k_z)]\operatorname{Re}[\Sigma^R_n(\Omega,k_z)]$ and $\operatorname{Im}[D^R(\Omega,q)]\operatorname{Re}[\Pi^R(\Omega,q)]$. In order to see that explicitly, one has to express $\operatorname{Re}[\Sigma^R_n(\Omega,k_z)]$ and $\operatorname{Re}[\Pi^R(\Omega,q)]$ in terms of the spectral functions by means of the following integral representation 
\begin{widetext}
\begin{align}\label{eq:spectral_Sigma}
 \Sigma_n^R(\Omega,k_z)&=g^2 \sum\limits_{\bar{n}}\int\frac{d^3q}{(2\pi)^3} X_{n\bar{n}}(q_\parallel)\int\limits_{-\infty}^{+\infty}\frac{dx d y}{\pi^2} \frac{\operatorname{Im}G_{\bar{n}}^R(x,k_z+q_z)\operatorname{Im}D^R(y,q)}{x+y-\Omega -i0^+} \Big(1+n_B(y)-n_F(x)\Big),\\ \label{eq:spectral_Pi}
 \Pi^R(y,q)&=\frac{m\omega_c g^2}{2\pi N}\sum\limits_{n\bar{n}}X_{n\bar{n}}(q_\parallel)\int\limits_{-\infty}^{+\infty}\frac{dk_z}{2\pi}\int\limits_{-\infty}^{+\infty}\frac{dx d \Omega}{\pi^2} \frac{\operatorname{Im}G_n^R(x,k_z+q_z)\operatorname{Im}G_{\bar{n}}^R(\Omega,k_z)}{x+y-\Omega+i0^+} \Big(n_F(\Omega)-n_F(x)\Big),
 \end{align}
which follows directly from analytic continuation of Eq.~\eqref{eq:Sigma_eq_new} and Eq.~\eqref{eq:Pi_eq}. After symmetrizing the integrals and making use of some identities involving the Fermi and Bose distributions, such as $ n_B(-y)=-1-n_B(y)$, $n_F(\Omega+y)=n_F(\Omega)n_B(y)/(1+n_B(y)-n_F(\Omega))$, as well as $\operatorname{Im}D^R(-y,q)=-\operatorname{Im}D^R(y,q)$, we obtain the following {\it exact} expression for the entropy 
\begin{equation} \label{eq:entropy_general_1}
\begin{aligned}
    S &=- \frac{m\omega_c}{2\pi^2}\int\limits_{-\infty}^{+\infty}d\Omega \frac{\partial n_F(\Omega)}{\partial T}\sum\limits_{n=0}^{+\infty}  \int\limits_{-\infty}^{+\infty}\frac{dk_z}{2\pi}\left\{\operatorname{Im}\ln\Big(-[G^R_n\left(\Omega,k_z\right)]^{-1}\Big) + \Re G^R_n(\Omega,k_z)\Im \Sigma^R_n(\Omega,k_z) \right\}
   \\
  &- \int\frac{d\Omega}{2\pi} \frac{\partial n_B(\Omega)}{\partial T} \int\frac{d^3q}{(2\pi)^3}\Big\{\operatorname{Im}\ln\left([D^R(\Omega,q)]^{-1}\right)+\operatorname{Im}\Pi^R(\Omega,q) \operatorname{Re}D^R(\Omega,q)\Big\}. 
  \end{aligned}
\end{equation}
\end{widetext}

We emphasize that Eq.~\eqref{eq:entropy_general_1} is formally exact in powers of $\omega_c/\mu$ and $g^2$ within the ME
approximation (i.e. if the vertex corrections are absent). It is worth mentioning that one can still express the entropy $S$ in terms of temperature derivatives acting only on the distributions even beyond the ME theory, but the simple structure in Eq.~\eqref{eq:entropy_general_1} will be then supplemented by additional terms \cite{vanderheyden1998self}. The crucial advantage of this representation is that the temperature derivatives of the distributions are sharply peaked around $\Omega= 0$, with the characteristic width of the order of $T$. Thus, in order to determine the leading order low-temperature behavior of the entropy it is sufficient to use the $T=0$ expressions for the self-energies. The explicit temperature dependence of the retarded self-energies  is always sub-leading since the effective range of frequencies contributing to the integrals in Eq.~\eqref{eq:entropy_general_1} already shrinks to zero at low temperatures. In other words, thermodynamic properties at the lowest temperatures are determined by the spectrum of
the elementary excitations near the Fermi surface (i.e. by the poles of the Green's function
$G_n^R(\Omega,k_z)$ as a function of $\ep_{nk_z}{-}\mu$ at $T=0$), see $\S$19.5 in \cite{AGD}.
In particular, this property demonstrates that Eq.~\eqref{eq:entropy_general_1} manifestly preserves the thermodynamic constraints in Eq.\eqref{eq:constraints}.

\section{Oscillations with undamped bosons}\label{sec:_z=1}
Using the formulation for the exact Migdal-Eliashberg entropy, Eq.~\eqref{eq:entropy_general_1}, we will now study the oscillatory entropy from which we will extract the DHVA magnetization.  While our main interest is in studying the case where the boson is overdamped with dynamic exponent $z=3$, we will study, in this section a somewhat artificial theory in which the boson remains undamped, while the fermion remains dressed into a marginal Fermi liquid.  Our reasons for studying this theory are two-fold.  First, it serves as a pedagogical ``warm-up" exercise, since the bosons remain free and the bosonic contribution to the entropy will remain negligible.  Moreover, the remaining fermionic contributions can be computed in a controlled fashion.  Second, it illustrates the importance of dynamical scaling laws and Landau damping: the DHVA magnetization extracted from this theory will be drastically different from that of the more physical case with an overdamped boson.   

Such undamped bosons can also arise from ME equations but with a key alteration in the bosonic sector: a small parameter $1/N$ ( with $N \rightarrow \infty$)  as a prefactor in the expression for $\Pi$:
\begin{equation}\label{eq:Pi_eq_N}
\begin{aligned}
   &\Pi(\omega_{\bar{m}},q)=  -\frac{g^2 T}{N}\sum\limits_m \hspace{-0.3em} \int\hspace{-0.3em}\frac{d^3k}{(2\pi)^3}\bar{G}(\omega_{\bar{m}}+\ep_m,k+q)\bar{G}(\ep_m,k)\\
   &=-\frac{g^2 T}{N}\sum\limits_{\substack{m\\n\bar{n}}}  \frac{X_{n\bar{n}} (q_\parallel)}{(2\pi l_B)^2} \hspace{-0.2em}\int\limits_{k_z} \hspace{-0.2em} G_n(k_z+q_z,\ep_m+\omega_{\bar{m}})G_{\bar{n}}(k_z,\ep_m).
   \end{aligned}
\end{equation}
The appropriate thermodynamic potential having the altered form of the saddle point equations is the same as before, but with a different pre-factor in the bosonic sector of $\Xi$: 
\begin{equation}
\begin{aligned}\label{eq:Xi_boson}
    \Xi_{\rm bos}&=\frac{NT}{2}\sum\limits_{m}\int\frac{d^3q}{(2\pi)^3}\ln\left([D_0(\omega_m,q)]^{-1}-\Pi(\omega_m,q)\right)\\
    &+\frac{NT}{2}\sum\limits_{m}\int\frac{d^3q}{(2\pi)^3}\Pi(\omega_m,q)D(\omega_m,q),
    \end{aligned}
\end{equation}
Such a theory can formally be derived in the case of a fermion belong to a fundamental representation of a $SU(N)$ global symmetry group, whereas the boson belongs to an adjoint representation of the group.  It describes intermediate scale behavior (i.e. at scales larger than $\mathcal O(1/N)$, where physical effects such as Landau damping set in) at a quantum critical point.  Although Landau damping is an $\mathcal O(1/N)$ effect, the fermions are dressed into a marginal Fermi liquid at leading order \cite{Mahajan2013,Fitzpatrick2014,Wang2017,Damia2019}. The critical point in this theory is reached by simply setting $m_b^2=0$ in the bare bosonic propagator.

From the modified thermodynamic potential above, one can repeat the steps of the previous section to obtain  the exact  expression for the ME entropy in this theory that differs from Eq.~\eqref{eq:entropy_general_1} only by an extra factor of $N$ in front of the bosonic contribution. Since the bosonic self-energy $\Pi^R$ is now suppressed by $1/N$ (cf. Eq.~\eqref{eq:Pi_eq_N}), then the term containing $\operatorname{Im}\ln\left([D^R(\Omega,q)]^{-1}\right)$ can be expanded up to the first order in $\Pi^R$. This gives 
\begin{widetext}
\begin{equation} \label{eq:entropy_undamped}
\begin{aligned}
    S &=- \frac{m\omega_c}{2\pi^2}\int\limits_{-\infty}^{+\infty}d\Omega \frac{\partial n_F(\Omega)}{\partial T}\sum\limits_{n=0}^{+\infty}  \int\limits_{-\infty}^{+\infty}\frac{dk_z}{2\pi}\left\{\operatorname{Im}\ln\Big(-[G^R_n\left(\Omega,k_z\right)]^{-1}\Big) + \Re G^R_n(\Omega,k_z)\Im \Sigma^R_n(\Omega,k_z) \right\}
   \\
  &+NS_{\rm free\; bos}
   + \int \frac{d\Omega}{2\pi} \frac{\partial n_B}{\partial T} \int\frac{d^3q}{(2\pi)^3}\operatorname{Im}D_0^R(\Omega,q)\operatorname{Re}\left[N\Pi^R(\Omega,q)\right].
  \end{aligned}
\end{equation}
\end{widetext}
Note that here the combination $N\Pi^R(\Omega,q)\sim \mathcal{O}(1)$ is evaluated with the full interacting fermionic Green's functions according to Eq.~\eqref{eq:spectral_Pi}, whereas the fermionic self-energy obeys the simplified self-consistency equation Eq.~\eqref{eq:spectral_Sigma} with $D^R(\Omega,q)$ replaced by the bare propagator $D^R_0(\Omega,q)$. In addition, $S_{\rm free\; bos}$ in Eq.~\eqref{eq:entropy_undamped} is the entropy of a single free bosonic mode (this term does not oscillate in a magnetic field, and thus we can ignore it).

Let us pause here and briefly summarize the remaining steps in the evaluation of \eqref{eq:entropy_undamped} that will be done below. First, we will single-out the leading order oscillatory contribution to the entropy assuming (and then verifying later) that the full fermionic self-energy is essentially independent of the Landau level $n$ in the regime of interest (i.e. at $T=0$ and for $\omega_c \ll \omega_D \ll E_F$). Next, we will argue that the contribution to the oscillations from the undamped bosonic modes is irrelevant at low temperatures, and thus, can be neglected. After that, we will evaluate the fermionic self-energy at $T=0$ including its oscillatory behavior as a function of frequency, and demonstrate that only the self-energy integrated over the perpendicular component of momentum $k_z$ retains a non-analytic form (i.e. marginal Fermi liquid features) at frequencies smaller than $\omega_c$, thus yielding a $T\ln T$ scaling of the entropy oscillations. In contrast, the non-nonanalyticity of the self-energy at the extremal orbit $k_z\approx 0$ is strongly smeared and no longer contributes to the low-temperature dependence of the entropy. Finally, we use the Maxwell relation 

\be\label{eq:Maxwell}
\frac{\partial S}{\partial B} = \frac{\partial M}{\partial T}
\ee
in order to extract the corresponding $T^2\ln T$ scaling of the magnetization $M$.

 Let us now carry out these steps explicitly. First, instead of following Luttinger's paradigm and expanding Eq.~\eqref{eq:entropy_undamped} in powers of the oscillatory self-energies, here we simply evaluate Eq.~\eqref{eq:entropy_undamped} by assuming that the full fermionic self-energy is essentially independent of the Landau level $n$ at $T=0$ and for $\omega_c \ll \omega_D \ll E_F$. Summing over Landau levels using the Poisson summation formula, and, whenever possible, evaluating the $k_z$-integrals in the stationary-phase approximation, we obtain
\begin{widetext}
\begin{equation} \label{eq:entropy_large2}
\begin{aligned}
    S_{\rm osc} &= \frac{(m\omega_c)^{3/2}}{4\pi^3} \sum\limits_{k=1}^{+\infty} \frac{(-1)^k}{k^{3/2}}   \int\limits_{-\infty}^{+\infty}d\Omega \frac{\partial n_F}{\partial T} e^{\frac{2\pi k}{\omega_c} \operatorname{Im}\Sigma^R(\Omega,0)}  \sin\left(\frac{2\pi  k} {\omega_c}\left[\Omega+\mu-\operatorname{Re}\Sigma^R(\Omega,0)\right]-\frac{\pi}{4}\right)\left(1-\frac{2\pi k}{\omega_c} \operatorname{Im}\Sigma^R(\Omega,0)\right)\\
    &-\frac{m}{4\pi^2} \int\limits_{-\infty}^{+\infty}d\Omega \frac{\partial n_F}{\partial T}\int\limits_{-\infty}^{+\infty}dk_z \operatorname{Re}\Sigma^R_{\rm osc}(\Omega,k_z)+\int\frac{d\Omega}{2\pi} \frac{\partial n_B}{\partial T} \int\frac{d^3q}{(2\pi)^3}\operatorname{Im}D_0^R(\Omega,q)\operatorname{Re}\left[N\Pi^R_{\rm osc}(\Omega,q)\right].
  \end{aligned}
\end{equation}
\end{widetext}
We emphasize that $\operatorname{Im}\Sigma^R(\Omega,0)$ appearing in the first line of Eq.~\eqref{eq:entropy_large2} contains both `smooth' and `oscillatory' contributions (i.e. this expression is beyond the usual Luttinger's expansion). Let us now briefly emphasize the origin of various terms in Eq.~\eqref{eq:entropy_large2}. The first line in this formula stems from the `oscillatory' part of the Poisson resummation of the first line in Eq.~\eqref{eq:entropy_undamped}. As a result, the $k_z$ integral is dominated by the extremal orbit at $k_z=0$ (the deviations from it are suppressed by extra powers of $\sqrt{\omega_c/\mu}$), and the rest of this expression depends on $\Sigma^R(\Omega,k_z=0)$ only. In contrast, the first term of the second line in Eq.~\eqref{eq:entropy_large2} originates from the `smooth' part of the Poisson resummation. Thus, the integral over $k_z$ is not confined to the extremal orbit yielding an `averaged' quantity $\int dk_z \operatorname{Re}\Sigma^R_{\rm osc}(\Omega,k_z)$. As we will see below, $\int dk_z \operatorname{Re}\Sigma^R_{\rm osc}(\Omega,k_z)$ and $\Sigma^R(\Omega,k_z=0)$ have very different behavior at frequencies below $\omega_c$. 

The computation of the contribution from the bosons (the last term in Eq.~\eqref{eq:entropy_large2} ) is the easiest, thus we proceed with obtaining the leading temperature dependence from this term first. This can be done by rescaling the frequencies as $\Omega=2 T x$ (where $x$ is a dimensionless integration variable $\sim 1$) and taking into account that the spectral function of the undamped boson has the following simple form
\be \label{eq:spectral_boson_z=1}
\operatorname{Im}D_0^R(\Omega,q)= \frac{\pi}{2c q}\Big[\delta(cq-\Omega)-\delta\left(cq+\Omega\right)\Big].
\ee
As a result, we find
\begin{equation}\label{eq:Pi_Ngg1_S}
    \frac{NT}{2\pi^2 c^3}\int\limits_0^{+\infty} \frac{ dx \;x^2}{\sinh^2x} \operatorname{Re}\Pi^R_{\rm osc}\left(2Tx, \frac{2Tx}{c}\right)\approx \frac{\pi^2 \bar{g}^2}{3}\beta \nu_{\rm osc} T.
\end{equation}
Also, in Eq.~\eqref{eq:Pi_Ngg1_S} we used $ \nu_{\rm osc}\equiv \frac{N}{g^2}\Pi^R_{\rm osc}(0,q\rightarrow 0)$ (this limit appears because for the typical frequency and momenta in Eq.~\eqref{eq:Pi_Ngg1_S} we have $\omega/v_Fq = 1/\beta \ll 1$). Note that $\nu_{\rm osc}$ is just the oscillatory part of the compressibility (see Appendix~\ref{sec:LD_B})
\begin{equation}\label{eq:compress22}
\begin{aligned}
    \nu_{\rm osc} &= \nu \sqrt{\frac{\omega_c}{2\mu}}\sum\limits_{k=1}\frac{(-1)^k}{\sqrt{k}}\cos\left(\frac{2\pi k \mu}{\omega_c}-\frac{\pi}{4}\right)\\
    &= \nu \sqrt{\frac{\omega_c}{2\mu}}\operatorname{Re}\left(e^{\frac{i\pi}{4}}\operatorname{Li}_{1/2}\left(-e^{-\frac{2i\pi \mu}{\omega_c}}\right)\right).
    \end{aligned}
\end{equation}
Note also that Eq.~\eqref{eq:Pi_Ngg1_S} contributes only at the linear order in temperature, and thus, it corresponds to just a weak renormalization of the free Fermi gas result.

On the other hand, the leading-temperature contribution stemming from the remaining (fermionic) terms in Eq.~\eqref{eq:entropy_large2} is a non-analytic function of temperature. In order to determine it, one has to first analyze the fermionic self-energies at $T=0$ including the oscillatory part, which we now present in the following subsection.

\subsection{Oscillatory part of the fermionic self-energy at $T=0$}\label{sec:secSEz1}
Our starting point is Eq.~\eqref{eq:spectral_Sigma} at $T=0$ - a self-consistency equation with the self-energy incorporated into the fermionic Green's functions, and with the bare bosonic spectral function $\operatorname{Im}D_0^R(\Omega,q)$, given in Eq.~\eqref{eq:spectral_boson_z=1}, instead of the full one $\operatorname{Im }D^R(\Omega,q)$ -- a simplification of the large-$N$ theory. Since the integral over $\Omega$ in Eq.~\eqref{eq:entropy_large2} is confined to the range $|\Omega|\ll T\ll \omega_c$, then we are only interested in the low-frequency asymptotic of $\Sigma^R_n(\Omega,k_z)$ for $|\Omega|\ll \omega_c$. Under these conditions, we solve the self-consistency equation iteratively, by first computing the r.h.s. of Eq.~\eqref{eq:spectral_Sigma} with the marginal Fermi liquid propagators \begin{equation}\label{eq:G_MF_n}
\begin{aligned}
    [\tilde{G}_{n}^R]^{-1} &= \Omega \left(1+\bar{g}^2 \ln\frac{\omega_D}{|\Omega|}\right)+ \frac{i \pi \bar{g}^2}{2}|\Omega|-\epsilon(n,k_z)+\mu. 
    \end{aligned}
\end{equation}
The delta function appearing in the r.h.s. of Eq.~\eqref{eq:spectral_boson_z=1} could be represented in the following form
\begin{equation}
\delta\left(\sqrt{q^2_\parallel+q_z^2}-\frac{\Omega}{c}\right)= \frac{\Omega \theta\left(\Omega-c |q_z|\right)}{\sqrt{\Omega^2-c^2q_z^2}} \delta\left(q_\parallel  -\sqrt{\frac{\Omega^2}{c^2}-q_z^2}\right)
\end{equation}
Next, we note that the form factor $ X_{nn'}(q_\parallel)$ in Eq.~\eqref{eq:A_nnn}, describing the matrix elements of $\exp\left(i \bm{q}_\parallel \bm{r}_\parallel\right)$ in the Landau basis, can be approximated as
\begin{equation}\label{eq:A_approx_appendix}
    X_{nn'}(q) \approx\frac{1}{\pi R_c q}\;,\quad 1/R_c\ll q\ll k_F.
\end{equation}
The upper limit on typical momentum transfer $q\ll k_F$ here allows one to replace Eq.~\eqref{eq:A_nnn} with the Bessel function (see Eq.~\eqref{eq:B_nnn}), whereas the lower limit $R_c^{-1}\ll q$ gurantees that only the large-argument asymptotic of that Bessel function matters, yielding Eq.~\eqref{eq:A_approx_appendix}. We emphasize that both of these conditions are well-satisfied for the typical in-plane momentum transfer on the r.h.s. of Eq.~\eqref{eq:spectral_Sigma} since for the undamped bosonic propagator we have $(q_\parallel)_{\rm typ} \sim \omega_c/c=\beta/R_c\gg 1/R_c$ and $v_F/c\gg 1$, but also at the same time $(q_\parallel)_{\rm typ}\ll k_F$ since $\omega_c\ll \omega_D$. Thus, under these circumstances, $X_{n\bar{n}}(q_\parallel)$ does not depend on $n$ and $\bar{n}$. As a consequence, $\Sigma_R^n(\Omega,k_z)\equiv \Sigma_R(\Omega,k_z)$ becomes essentially independent of the Landau level $n$, justifying our simple expression for the entropy Eq.~\eqref{eq:entropy_large2}.

With this approximation, we first compute the imaginary part of the self-energy. After making use of the Poisson summation formula, and neglecting higher order corrections of the order of $\omega_D/\mu \ll 1$, we obtain the following oscillatory contribution for $k_z=0$ and for the average over all $k_z$ 
\begin{widetext}
\begin{equation}\label{eq:Sigma_Im_cal_z=1}
    \begin{aligned}
&\operatorname{Im}\Sigma^R_{\rm osc}(\Omega,0) \hspace{-0.1em}=\hspace{-0.1em} -\pi\bar{g}^2 \hspace{-0.2em}\int\limits_{0}^{|\Omega|} \hspace{-0.2em}d\ep \sum\limits_{k=1}^{+\infty}(-1)^k e^{-\frac{\pi^2\bar{g}^2 k}{\omega_c}|\ep|}J_0\left(\frac{\pi k (|\Omega|\hspace{-0.2em}-\hspace{-0.2em}\ep)^2}{2m c^2\omega_c}\right)\hspace{-0.1em}\cos\left(\frac{2\pi k}{\omega_c}\left(\mu+\operatorname{sgn}(\Omega)\ep\hspace{-0.2em}\left[1\hspace{-0.1em}+\hspace{-0.1em}\bar{g}^2\ln\frac{\omega_D}{|\ep|}\right]\right)-\hspace{-0.2em}\frac{\pi k (|\Omega|\hspace{-0.2em}-\hspace{-0.2em}\ep)^2}{2m c^2\omega_c}\right),\\
&\int\limits_{-\infty}^{+\infty}dk_z\operatorname{Im}\Sigma^R_{\rm osc}(\Omega,k_z) = -\frac{\pi\bar{g}^2 k_F}{\sqrt{2}}  \sqrt{\frac{\omega_c}{\mu}} \int\limits_{0}^{|\Omega|} \hspace{-0.2em}d\ep  \sum\limits_{k=1}^{+\infty}\frac{(-1)^k }{\sqrt{k}} e^{-\frac{\pi^2\bar{g}^2 k}{\omega_c}|\ep|}\cos\left(\frac{2\pi k}{\omega_c}\left(\mu+\operatorname{sgn}(\Omega)\ep\left[1+\bar{g}^2\ln\frac{\omega_D}{|\ep|}\right]\right)-\frac{\pi}{4}\right).
\end{aligned}
\end{equation}
\end{widetext}
In the first line in Eq.~\eqref{eq:Sigma_Im_cal_z=1}, the integral over $q_z$ was performed by means of the following identity
\begin{equation}
    \int\limits_{-a}^{+a}\frac{dt}{\sqrt{a^2-t^2}}\cos\left(b-ct^2\right)=\pi J_0\left(\frac{a^2c}{2}\right)\cos\left(b- \frac{a^2c}{2}\right), 
\end{equation}
where $J_0(x)$ is the Bessel function. Note that this Bessel function cannot be simply replaced by its large argument asymptotic in the naive limit $\omega_c\rightarrow 0$ (this limit corresponds to the saddle point approximation of the original $q_z$-integral) because the remaining integral over frequency would be logarithmically divergent. In contrast, the frequency integral in the second line of Eq.~\eqref{eq:Sigma_Im_cal_z=1} is well-behaved even after the $q_z$-integral is carried out in the saddle point limit.

\begin{figure*}[t!]
\centerline{\includegraphics[width=1.0\textwidth]{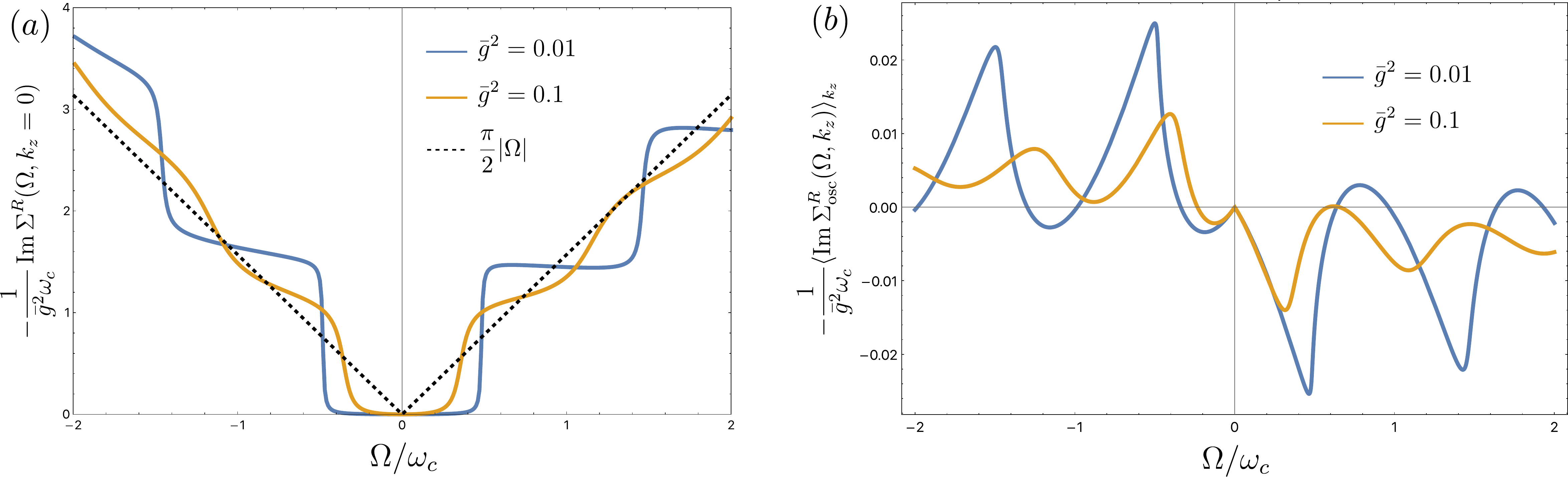}}
\caption{(a) Imaginary part of the full self-energy (`smooth' and `oscillatory' parts combined) evaluated at the extremal orbit $k_z=0$. The dashed line indicates the marginal FL self-energy at $B=0$. (b) Imaginary part of the `oscillatory' contribution to the self-energy averaged over all values of $k_z$. Here $\langle...\rangle_{k_z}$ stands for $\int dk_z  .../2k_F$. In both (a) and (b), the parameters are $\omega_c/\mu=200$ and $\omega_D/\omega_c=20$, with two different values of the coupling constant $\bar{g}^2$. }
\label{fig:z=1_SE}
\end{figure*}

Our next step is to estimate the low-frequency behavior of the imaginary part of the full self-energy evaluated at $k_z=0$, i.e. $\operatorname{Im}\Sigma^R(\Omega,0)$ for $|\Omega|\ll \omega_c$.
 To this end, we perform the rescaling $\ep = |\Omega|t$ in Eq.~\eqref{eq:Sigma_Im_cal_z=1}, and expand in $|\Omega|/\omega_c\ll 1$. As a result, we obtain
\begin{equation}\label{eq:Im_osc_z=1}
\begin{aligned}
\operatorname{Im}\Sigma^R_{\rm osc}(\Omega,0)\approx -\pi \bar{g}^2|\Omega|\sum\limits_{k=1}^{+\infty}(-1)^k\cos\left(\frac{2\pi k\mu}{\omega_c}\right)=\frac{\pi\bar{g}^2}{2}|\Omega|
    \end{aligned}
\end{equation}
Here the series is assumed to be regularized by an extra factor $\exp\{-k 0^+\}$. A few comments are in order. First, we emphasize that, despite its appearance, Eq.~\eqref{eq:Im_osc_z=1} contains no assumptions about the strength of the coupling - higher order dependence on $\bar{g}^2$ enters only via the irrelevant corrections to Eq.~\eqref{eq:Im_osc_z=1} which we ignore for now. Second, even though the sum over all harmonics in Eq.~\eqref{eq:Im_osc_z=1} yields a formally non-oscillating result at the lowest frequencies $|\Omega|\ll \omega_c$, we stress that higher-order contributions beyond this asymptotic limit still oscillate both as a function of $\mu/\omega_c$ and $\Omega/\omega_c$ (this can be seen from numerical evaluation of $\operatorname{Im}\Sigma^R(\Omega,0)$, see Fig.~\ref{fig:z=1_SE}).
Next, most crucially, Eq.~\eqref{eq:Im_osc_z=1} coincides with the `smooth' part of $\operatorname{Im}\Sigma^R(\Omega,0)$, but has the opposite sign. Thus, the $|\Omega|$-linear term cancels out, and higher powers of $|\Omega|$ must be retained. Stated differently, this cancellation between the two parts of Poisson resummation (`smooth' and `oscillatory') implies that the total $\operatorname{Im}\Sigma^R(\Omega,0)$ vanishes faster than $|\Omega|$ at small $|\Omega|\ll \omega_c$. Numerical evaluation of $\operatorname{Im}\Sigma^R(\Omega,0)$ at several values of $\bar{g}^2$ is shown in Fig.~\ref{fig:z=1_SE}, from which one can see that the marginal Fermi liquid effects are strongly suppressed by magnetic field for the excitations at the extremal orbit $k_z=0$. This result is reminiscent of the behavior observed in {\it purely two-dimensional} NFL models in a magnetic field \cite{StampSE,Thompson2010,QO_random_Yukawa}.

The real part of the self-energy can be obtained via the Kramers-Kronig relation
\begin{eqnarray}
\operatorname{Re}\Sigma^R(\Omega,0)= \frac{1}{\pi}\int\limits_{-\omega_D}^{\omega_D}\frac{d\omega }{\omega-\Omega} \operatorname{Im}\Sigma^R(\omega,0),
\end{eqnarray}
where the integral is understood in the principle value sense. First, we note that the value at $\Omega=0$ remains finite due to the fact that $\operatorname{Im}\Sigma^R(\Omega,0)$ is not an even function of frequency (see Fig.~\ref{fig:z=1_SE}). However, this oscillating temperature-independent constant corresponds to a shift of the chemical potential, and thus, has no significance. The lowest order frequency dependent term could be computed by looking at the odd part  
\begin{equation}
[\operatorname{Re}\Sigma^R(\Omega,0)]_{\rm odd} = \frac{\Omega}{\pi}\hspace{-0.3em}\int\limits_{-\omega_D}^{\omega_D}\hspace{-0.3em}\frac{d\omega \operatorname{Im}\Sigma^R(\omega,0)}{\omega^2-\Omega^2}\;. 
\end{equation}
The crucial point here is that one can now safely set $\Omega{=}0$ in the integrand because  $\operatorname{Im}\Sigma^R(\omega,0)$ vanishes faster than $|\omega|$ at $|\omega|\ll \omega_c$ rendering the integral convergent. This implies that the total real part of the self-energy has the following low-energy expansion
\begin{equation}\label{eq:ReSigma_z=1_W}
\operatorname{Re}\Sigma^R_{}(\Omega,0) \approx \text{const} -\bar{g}^2 \Omega \ln \left(\frac{\omega_D}{\omega_c}\right)+\mathcal{O}(\Omega),
\end{equation}
We emphasize that the logarithm in the second term is cut-off at $\omega_c$ instead of $|\Omega|$ (as it happens in the absence of magnetic field) due to the smearing of non-analyticity in $\operatorname{Im}\Sigma^R(\Omega,0)$ at low frequencies. This result, once more, illustrates that the excitations near the extremal orbit $k_z=0$ remain coherent at lowest scales in the presence of a magnetic field.

Next, we deal with the second line in Eq.~\eqref{eq:Sigma_Im_cal_z=1}, representing the self-energy averaged over all values of $k_z$. Although we are still only interested in the low-frequency limit, i.e. $|\Omega|\ll \omega_c$, we note that the perturbative self-energy at the leading order in $\bar{g}^2$ can be computed exactly even for a much broader range of frequencies $|\Omega| \ll \omega_D$ 
\begin{equation}
\label{eq:Im_Sigma_pert_k}
\begin{aligned}
     &\int\limits_{-\infty}^{+\infty}dk_z\operatorname{Im}\Sigma^R_{\rm osc}(\Omega,k_z)\approx -\bar{g}^2m^{1/2}\omega_c^{3/2}\sum\limits_{k=1}^{+\infty}\frac{(-1)^k}{k^{3/2}}\\
     &\times \sin\left(\frac{\pi k |\Omega|}{\omega_c}\right)\cos\left(\frac{2\pi k}{\omega_c}\left[\mu+\frac{\Omega}{2}\right]-\frac{\pi}{4}\right)+\mathcal{O}(\bar{g}^4).
     \end{aligned}
\end{equation}
If now, in addition, the low-frequency limit is taken, then the expression above results in a linear-in-$|\Omega|$ scaling, whereas the omitted terms $\mathcal{O}(\bar{g}^4)$ contribute only to the higher powers of $|\Omega|$ (i.e. they are irrelevant, similarly to the situation with Eq.~\eqref{eq:Im_osc_z=1}). Therefore, even for a finite value of $\bar{g}^2$, the leading asymptotic for $|\Omega|\ll \omega_c$ is given by 
\begin{equation}\label{eq:ImSigmakzaveragedz=1}
     \int\limits_{-\infty}^{+\infty}dk_z\operatorname{Im}\Sigma^R_{\rm osc}(\Omega,k_z)\approx  - \pi k_F \bar{g}^2\frac{\nu_{\rm osc} }{\nu }|\Omega|,
\end{equation}
irrespective of the presence of the  marginal Fermi liquid self-energy in the r.h.s. of Eq.~\eqref{eq:Sigma_Im_cal_z=1}. Numerical evaluation of $\int dk_z\operatorname{Im}\Sigma^R_{\rm osc}(\Omega,k_z)$ is shown in Fig.~\ref{fig:z=1_SE}.



These expressions allow us to compute the integral over $k_z$ of the real part of the self-energy by means of the Kramers-Kronig relations. We find
\begin{equation}\label{eq:Sigma_k_z_int}
\begin{aligned}
  \int\limits_{-\infty}^{+\infty}dk_z\operatorname{Re}\Sigma^R_{\rm osc}(\Omega,k_z)&=\int\limits_{-\omega_D}^{\omega_D}\frac{d\omega}{\pi} \int\limits_{-\infty}^{+\infty} dk_z\frac{ \operatorname{Im}\Sigma^R_{\rm osc}({\omega},k_z)}{\omega-\Omega} \\
  &\approx  - 2 k_F \bar{g}^2\frac{\nu_{\rm osc} }{\nu }\Omega\ln\frac{\omega_c}{|\Omega|}
  \end{aligned}
\end{equation}
at low frequencies $|\Omega|\ll \omega_c \ll \omega_D$. Note that the limits of integration over $\omega$ could be extended to infinity - the convergence at large frequencies is guaranteed by the oscillatory behavior of the self-energy as a function of $\omega$ (for instance, note the factor $\sin(\pi k |\Omega|/\omega_c)$ in Eq.~\eqref{eq:Im_Sigma_pert_k}).

It is also worth mentioning that the same results for the self-energy can be obtained on the Matsubara axis at $T=0$. For instance, one can start with Eq.~\eqref{eq:Sigma_eq_new} with $D_0(\omega_m,q)$ instead of $D(\omega_m,q)$, make use of the approximation in Eq.~\eqref{eq:A_approx_appendix}, and then perform the Poisson resummation. The result at $k_z=0$ reads as
\begin{equation}
\begin{aligned}
    \Sigma^{}_{\rm osc}(\ep,0) &\approx \frac{\bar{g}^2}{2}\sum\limits_{k=1}^{\infty} \int d\bar{\ep}\operatorname{sgn}(\bar{\ep})K_0\left(\frac{i\pi k (\ep-\bar{\ep})^2}{2c^2m\omega_c}\operatorname{sgn}(\bar{\ep})\right)\\
    & \times \exp\left\{-\frac{2\pi k }{\omega_c}|\bar{\ep}|\left(1+\bar{g}^2\ln\frac{\omega_D}{|\bar{\ep}|}\right)\right.\\&+\left.\frac{2\pi i k}{\omega_c}\left(\mu+\frac{(\ep-\bar{\ep})^2}{4mc^2}\right)\operatorname{sgn}(\bar{\ep})\right\}.
    \end{aligned}
\end{equation}
where we replaced the Matsubara sum with the continuous integral at $T=0$, and $K_0(x)$ is the modified Bessel function of the second kind. After taking the low-$\ep$ limit, we obtain 
\begin{equation}
\Sigma^{}_{\rm osc}(\ep,0)\approx \text{const} -\bar{g}^2 \ep \ln \frac{ \omega_c}{ |\ep|}+\mathcal{O}(\ep),
\end{equation}
where, once again, the sum over all harmonics resulted in a formally non-oscillating contribution that cancels the `smooth' part $\Sigma^{}(\ep,0)=\bar{g}^2 \ep \ln \frac{ \omega_c}{ |\ep|}$ leaving behind only higher order terms, in agreement with Eq.~\eqref{eq:ReSigma_z=1_W}.

In principle, one can now substitute these one-loop oscillatory self-energies back into the r.h.s. of Eq.~\eqref{eq:Sigma_Im_cal_z=1} in order to check if it affects the result in any way. However, since the low frequency asymptotics that we found turned out to be completely insensitive to the presence of the self-energies, then it is clear that Eq.\eqref{eq:ReSigma_z=1_W} and Eq.\eqref{eq:Sigma_k_z_int} will remain fully self-consistent upon further iterations of Eq.~\eqref{eq:Sigma_Im_cal_z=1}. 

\subsection{Oscillations of the entropy and magnetization}
The analysis of the self-energies performed in the previous section now allows us to evaluate the leading order temperature scaling of the entropy in Eq.~\eqref{eq:entropy_large2}. As was already mentioned before, the bosonic part only yields a $T$-linear contribution which is of no interest to us. Moreover, the absence of any non-analyticities in $\Sigma^R(\Omega,0)$ at $ |\Omega|\ll \omega_c$ results in a trivial $T$-linear scaling of the first term in Eq.~\eqref{eq:entropy_large2} as well. Finally, it turns out that the main contribution stems from the second term in \eqref{eq:entropy_large2} involving the the self-energy averaged over all $k_z$. This self-energy still retains a marginal Fermi liquid form with some oscillatory pre-factor, see Eq.~\eqref{eq:Sigma_k_z_int}. As a result, we obtain the following expression
\begin{equation}\label{eq:S_z=1_TlnT}
     S_{\rm osc} \approx  \frac{\pi^2 \bar{g}^2}{3} \nu_{\rm osc} T  \ln \frac{\omega_c}{ T} +\mathcal{O}(T).
\end{equation}
This expression has a clear physical interpretation: in the absence of the field, the entropy is given by $S(B=0)=\frac{\pi^2 \bar{g}^2}{3} \nu T \ln \frac{\omega_D}{ T}$, and thus, Eq.~\eqref{eq:S_z=1_TlnT} simply captures the fact that the compressibility $\nu$ oscillates in a magnetic field.

Finally, we can make use of the Maxwell relation Eq.~\eqref{eq:Maxwell}, and obtain the low-temperature behavior of the DHVA magnetization
\begin{equation}\label{eq:DHVA_z=1}
\begin{aligned}
    M_{\rm osc}&=\int\limits_{0}^{T}dT \frac{\partial S_{\rm osc}}{\partial B}\approx M_{\rm osc}(T=0)\\&+\frac{\pi^2 \bar{g}^2}{6} \frac{d\nu_{\rm osc} }{dB}T^2  \ln \frac{\omega_c}{ T}+\mathcal{O}(T^2).
    \end{aligned}
\end{equation}
The $\sim T^2\ln T$ scaling of the magnetization, together with the expression for the entropy Eq.~\eqref{eq:S_z=1_TlnT}, are the central results of this section. In particular, they imply that the oscillatory part of the thermodynamic potential acquires the same form as in  Eq.~\eqref{eq:Xi_0_osc_form}, with the amplitude $A_k(T)$ given by
\be\label{eq:A_z=1_final}
A_k(T)-A_k(T=0)\approx -\frac{\pi^2\bar{g}^2 k}{6 \omega_c} T^2\ln\frac{\omega_c}{T}.
\ee
for $T\ll \omega_c$. We will discuss these results in Sec.~\ref{sec:Discussion} in more detail.

\section{Effect of Landau damping on DHVA oscillations}
\label{sec:_z=3}
In this section we return to the full ME problem involving both fermionic and bosonic degrees of freedom on equal footing, as described by a system of two coupled equations \eqref{eq:spectral_Sigma} and \eqref{eq:spectral_Pi}. The derivation leading to Eq.~\eqref{eq:entropy_large2} remains essentially unchanged, with the only minor modification concerning the bosonic contribution. The full expression now reads as follows
\begin{widetext}
\begin{equation} \label{eq:entropy_large}
\begin{aligned}
    S_{\rm osc} &= \frac{(m\omega_c)^{3/2}}{4\pi^3} \sum\limits_{k=1}^{+\infty} \frac{(-1)^k}{k^{3/2}}   \int\limits_{-\infty}^{+\infty}d\Omega \frac{\partial n_F}{\partial T} e^{\frac{2\pi k}{\omega_c} \operatorname{Im}\Sigma^R(\Omega,0)}  \sin\left(\frac{2\pi  k} {\omega_c}\left[\Omega+\mu-\operatorname{Re}\Sigma^R(\Omega,0)\right]-\frac{\pi}{4}\right)\left(1-\frac{2\pi k}{\omega_c} \operatorname{Im}\Sigma^R(\Omega,0)\right)\\
    &-\frac{m}{4\pi^2} \int\limits_{-\infty}^{+\infty}d\Omega \frac{\partial n_F}{\partial T}\int\limits_{-\infty}^{+\infty}dk_z \operatorname{Re}\Sigma^R_{\rm osc}(\Omega,k_z)- \int\frac{d\Omega}{2\pi} \frac{\partial n_B}{\partial T} \hspace{-0.3em}\int \hspace{-0.3em}\frac{d^3q}{(2\pi)^3}\Big\{\operatorname{Im}\ln\left([D^R(\Omega,q)]^{-1}\right)+\operatorname{Im}\Pi^R(\Omega,q) \operatorname{Re}D^R(\Omega,q)\Big\}
  \end{aligned}
\end{equation}
\end{widetext}
As a first step, we consider the bosonic contribution. We recall that the spectral function of the Landau overdamped boson in the absence of the field is given by (see Appendix \ref{sec:self-energy} for details)
\be \label{eq:z=3_dynamical_scaling}
\operatorname{Im}\tilde{D}^R(\Omega,q)= \frac{\gamma \Omega q}{c^4 q^6 +\gamma^2 \Omega^2},\quad  \operatorname{Im}\tilde{\Pi}^R(\Omega,q)= \frac{\gamma \Omega}{q},
\ee
where we defined $\gamma\equiv \pi g^2\nu/2v_F$. Here we already assumed that the boson is tuned to criticality by setting the bare mass to be $m_b^2=\nu g^2$ in order to cancel the static part of $\tilde{\Pi}^R(\Omega,q)$ (see the first term in Eq.~\eqref{eq:Pi4}). In order to single out the oscillatory contribution stemming from $\Pi^R_{\rm osc}(\Omega,q)$ it is useful to add and subtract the same expression but with $\Pi^R_{\rm osc}(\Omega,q)$ set to zero. This leads to  
\begin{equation}\label{eq:z=3_boson_cont}
    \begin{aligned}
        S_{\rm osc}^{(bos)}=&\int\hspace{-0.2em}\frac{d\Omega}{2\pi} \frac{\partial n_B}{\partial T} \hspace{-0.3em}\int \hspace{-0.3em}\frac{d^3q}{(2\pi)^3}\Big\{\operatorname{Im}\ln\frac{1}{1-\tilde{D}^R(\Omega,q) \Pi_{\rm osc}^R(\Omega,q)}\\
        & \left.-\operatorname{Im}\tilde{\Pi}^R(\Omega,q) \operatorname{Re}\frac{[\tilde{D}^R(\Omega,q)]^2 \Pi_{\rm osc}^R(\Omega,q)}{1-\tilde{D}^R(\Omega,q) \Pi_{\rm osc}^R(\Omega,q)}\right\}.
    \end{aligned}
\end{equation}
Moreover, since we are only interested in the leading order $\mathcal{O}(\sqrt{\omega_c/\mu})$ contribution to the oscillations, it is sufficient to expand Eq.~\eqref{eq:z=3_boson_cont} up to the first power of $\Pi_{\rm osc}^R(\Omega,q)$ (as we show in Appendix \ref{sec:LD_B}, the oscillatory part of the bosonic self-energy always remains a small correction. This is in contrast to the fermionic self-energy that could be strongly enhanced at low frequencies, cf. Eq.~\eqref{eq:Im_osc_z=1}). We obtain
\begin{equation}\label{eq:z=3_boson_cont_2}
    \begin{aligned}
         S_{\rm osc}^{(bos)}\approx &\int\frac{d\Omega}{2\pi} \frac{\partial n_B}{\partial T} \int\hspace{-0.2em}\frac{d^3q}{(2\pi)^3}\hspace{-0.1em}\left\{\operatorname{Im}\tilde{D}^R(\Omega,q)\operatorname{Re}\Pi_{\rm osc}^R(\Omega,q)\right.\\
        &-\left.\operatorname{Im}\tilde{\Pi}_{}^R(\Omega,q)\operatorname{Re}\left([\tilde{D}^R(\Omega,q)]^2\Pi_{\rm osc}^R(\Omega,q)\right)\right\}.
    \end{aligned}
\end{equation}

The low-temperature behavior of this expression can be estimated as follows. As in the previous section, we rescale frequencies as $\Omega=2Tx$, where $x$ is a dimensionless variable of the order of one. This eliminates any temperature dependence from the derivative of the distribution function. Next, one can make use of the $z=3$ dynamical scaling (see Eq.~\eqref{eq:z=3_dynamical_scaling}), and rescale all momenta as $q\rightarrow T^{1/3}y$. In this case, the oscillatory part of the bosonic self-energy becomes $\Pi_{\rm osc}(2Tx, T^{1/3}y)$. In the $T\rightarrow 0$ limit, this quantity approaches the oscillatory part of the compressibility $\Pi_{\rm osc}(2Tx, T^{1/3}y)\approx g^2 \nu_{\rm osc}$ since in this regime we have $\omega/v_Fq \sim T^{2/3}\rightarrow 0$. The remaining integrals over $y$ and $x$ can be evaluated exactly and yield
\begin{equation}\label{eq:S_4_z=3}
     S_{\rm osc}^{(bos)} \approx   \frac{8c_1\pi^{1/3}}{9\sqrt{3}} \bar{g}^{8/3} (\beta \omega_D)^{2/3}\nu_{\rm osc}  T^{1/3},
\end{equation}
where $c_1=\int_{0}^{+\infty}dx\; x^{4/3}/\sinh^2x \approx 3.4$.

Next, we turn to the evaluation of the fermionic contribution to Eq.~\eqref{eq:entropy_large}. As in the undamped $z=1$ case, we first need to consider the low-frequency limit of the fermionic self-energy.

\subsection{Oscillatory part of the fermionic self-energy at $T=0$}\label{sec:dim_red_sec}
As in Sec.~\ref{sec:_z=1}, we begin by evaluating Eq.~\eqref{eq:spectral_Sigma} iteratively while neglecting the oscillatory parts of the self-energy in the r.h.s. of this equation. At the same time, we retain the marginal FL self-energy and the Landau damping term for the bosons. The latter only improves the accuracy of our approximation in Eq.~\eqref{eq:A_approx_appendix} since the dynamical exponent $z=3$ implies the typical transferred momentum $q_{\rm typ} \sim k_F^{2/3}R_c^{-1/3} \gg R_c^{-1}$. The resulting expression for the imaginary part of the self-energy has the following form
\begin{widetext}
\begin{equation}\label{eq:z=3_Sigma_Im}
    \begin{aligned}
    &\operatorname{Im}\Sigma_{\rm osc}^R(\Omega,0) = -\frac{2^{5/6}\bar{g}^{4/3} }{3 \sqrt{3} \pi^{1/3}} (\beta\omega_D)^{1/3}\sqrt{\frac{\omega_c}{\mu}}\sum\limits_{k}\frac{(-1)^k}{\sqrt{k}}\int\limits_0^{|\Omega|}d\ep \frac{  e^{-\frac{\pi^2\bar{g}^2 k}{3\omega_c}\ep}}{(|\Omega|-\ep)^{1/3}}\cos\left(\frac{2\pi k }{\omega_c}\hspace{-0.1em}\left(\mu+\operatorname{sgn}(\Omega) \ep \left[1+\frac{\bar{g}^2}{3}\ln\frac{\Lambda_D}{\ep}\right]\right)\hspace{-0.1em}-\hspace{-0.1em}\frac{\pi}{4}\right)\\
        &\int\limits_{-\infty}^{+\infty}dk_z\operatorname{Im}\Sigma_{\rm osc}^R(\Omega,k_z)=-\frac{\pi \bar{g}^2 k_F}{3\sqrt{2}}\sqrt{\frac{\omega_c}{ \mu}}\sum\limits_{k=1}^{+\infty}\frac{(-1)^k}{\sqrt{k}} \int\limits_0^{|\Omega|} d\ep   e^{-\frac{\pi^2\bar{g}^2 k}{3\omega_c}\ep}\cos\left(\frac{2\pi k }{\omega_c}\left(\mu+\operatorname{sgn}(\Omega)\ep \left[1+\frac{\bar{g}^2}{3}\ln\frac{\Lambda_D}{\ep}\right]\right)-\frac{\pi}{4}\right). 
    \end{aligned}
    \end{equation}

\begin{figure*}[t!]
\centerline{\includegraphics[width=1.0\textwidth]{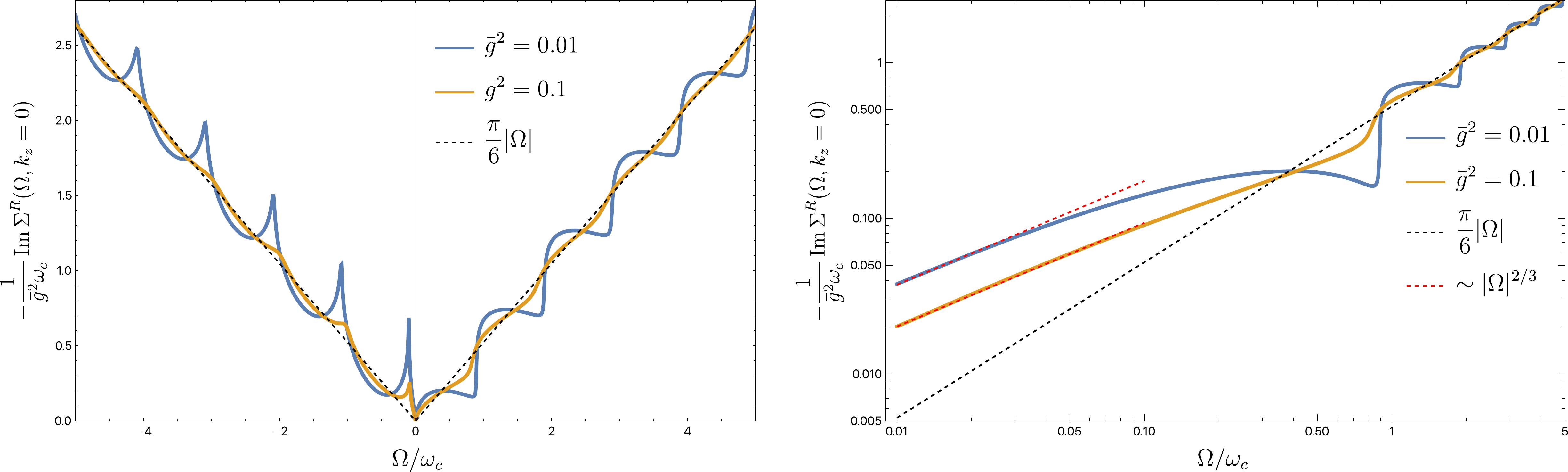}}
\caption{(Left): Numerical evaluation of the imaginary part of the full self-energy at extremal orbit $k_z=0$ in the overdamped case, based on Eq.~\eqref{eq:z=3_Sigma_Im}, with $\mu/\omega_c=200.6$, $\Lambda_D/\omega_c=10$. (Right): The same but on a log-log scale, with emphasized asymptotic $\sim |\Omega|^{2/3}$ behavior for $|\Omega|\ll \omega_c$ (cf. Eq.~\eqref{eq:Im_Sigma_k0z3}).}
\label{fig:Sigmaz=3}
\end{figure*}
It is instructive to compare these equations with the $z=1$ undamped case, Eq.~\eqref{eq:Sigma_Im_cal_z=1}. The self-energy at the extremal orbit $k_z=0$ acquires an additional small prefactor $\sqrt{\omega_c/\mu}$ stemming from the limited phase volume available in the saddle point approximation of the intermediate $k_z$ momentum integral. At the same time, the self-energy averaged over all $k_z$ has the same structure as in the undamped case Eq.~\eqref{eq:Sigma_Im_cal_z=1}, up to a trivial replacement $\bar{g}^2\rightarrow \bar{g}^2/3$. Numerical evaluation of the full self-energy $\operatorname{Im}\Sigma^R(\Omega,k_z=0)$, including the oscillatory part Eq.~\eqref{eq:z=3_Sigma_Im}, is depicted in Fig.~\ref{fig:Sigmaz=3} for several values of the coupling constant $\bar{g}^2$. At the leading order in $\bar{g}^2$, the integral over frequency $\ep$ in Eq.~\eqref{eq:z=3_Sigma_Im} can be computed analytically, and we obtain
\begin{equation}
    \begin{aligned}
    \operatorname{Im}\Sigma_{\rm osc}^R(\Omega,0)&\approx -\frac{\bar{g}^{4/3} (\beta\omega_D)^{1/3} }{2^{1/6} \sqrt{3} \pi^{1/3}}  \sqrt{\frac{\omega_c}{\mu}}|\Omega|^{2/3}\sum\limits_{k}\frac{(-1)^k}{\sqrt{k}} \left[\cos\left(\frac{2\pi k \mu}{\omega_c}-\frac{\pi}{4}\right) \hspace{-0.2em}\mathbin{_1 F_2}\left(1;\frac{5}{6},\frac{4}{3};
    -\pi^2 k^2 \frac{\Omega^2}{\omega_c^2}\right)\right.
  \\& \left.- \frac{6\pi k \Omega}{5\omega_c}\sin\left(\frac{2\pi k \mu}{\omega_c}-\frac{\pi}{4}\right) \hspace{-0.2em}\mathbin{_1 F_2}\left(1;\frac{4}{3},\frac{11}{6};
    -\pi^2 k^2 \frac{\Omega^2}{\omega_c^2}\right)\right]+\mathcal{O}(\bar{g}^4).
    \end{aligned}
\end{equation}
\end{widetext}
where $\mathbin{_1 F_2}$ is the hypergeometric function.  After taking the low-frequency limit $|\Omega|\ll \omega_c$, we find
\begin{equation}\label{eq:Im_Sigma_k0z3}
    \operatorname{Im}\Sigma_{\rm osc}^R(\Omega,0)\approx -\frac{2^{1/3} \bar{g}^{4/3}}{\sqrt{3}\pi^{1/3}} (\beta \omega_D)^{1/3} |\Omega|^{2/3} \frac{\nu_{\rm osc}}{\nu}.
\end{equation}

\begin{figure}[h!]
    \centering
    \includegraphics[width=0.99\columnwidth]{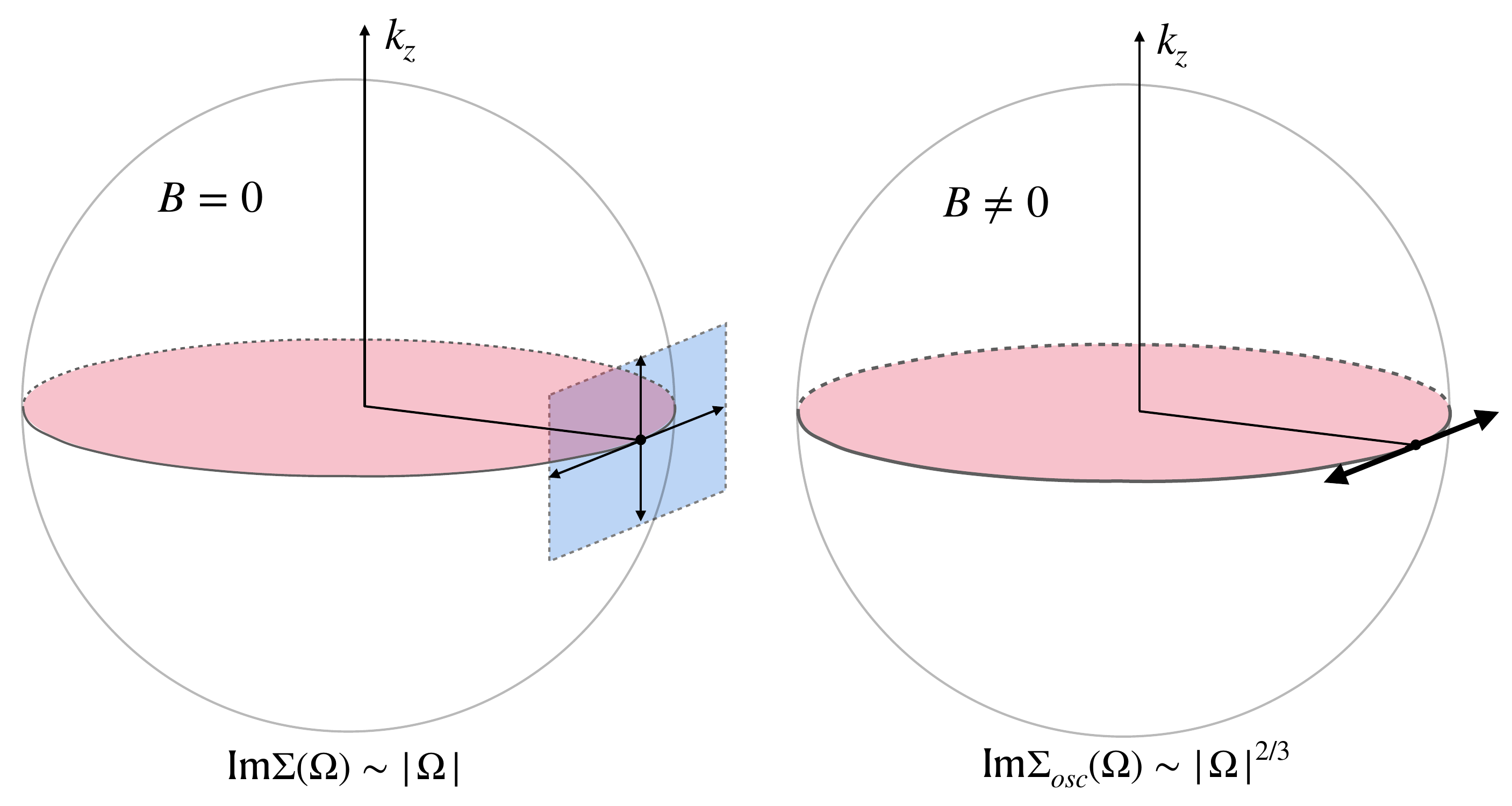}
    \caption{(Left): In the ME theory at $B=0$, typical transferred momentum is predominantly transverse to the Fermi surface due to the $z=3$ overdamped dynamics of the bosons. This leads to a $\sim|\Omega|$ scaling of the imaginary part of the self-energy for all momentum states on the Fermi surface. (Right): At finite $B$, only the states in the width $\sim \sqrt{\omega_c/\mu}$ around the extremal orbit $k_z= 0$ contribute to the oscillations. Thus, the oscillatory part of the self-energy at the external momenta $k_z=0$ is contributed only by the scattering processes that do not take the fermion away from $k_z\approx 0$ in the intermediate state. This condition makes the bosonic kinematics essentially two-dimensional, leading to the $|\Omega|^{2/3}$ scaling.}
    \label{fig:dim_red}
\end{figure}

Remarkably, this expression exhibits a $\sim|\Omega|^{2/3}$ scaling which is a well-known characteristic feature of a NFL self-energy emerging from the zero-field ME equations \eqref{selfconsistentzerofield} but in {\it{two spatial dimensions}}. Intuitively, this `dimensional reduction' can be understood from Fig.~\eqref{fig:dim_red} emphasizing a reduction of the phase volume available for scattering processes that keep the fermion on the extremal orbit $k_z=0$, rendering the boson's kinematics effectively two-dimensional. We also note that the presence of the marginal self-energy in the r.h.s. of Eq.~\eqref{eq:z=3_Sigma_Im} is not essential for the resulting lowest-frequency $\sim|\Omega|^{2/3}$ asymptotic in Eq.~\eqref{eq:Im_Sigma_k0z3} (in fact, Eq.~\eqref{eq:Im_Sigma_k0z3} is self-consistent upon further iterations of Eq.~\eqref{eq:spectral_Sigma}).

The expression for the $k_z-$averaged self-energy is identical to our previous result in Eq.~\eqref{eq:Im_Sigma_pert_k}, up to an additional prefactor $1/3$. Thus, in the low frequency limit, we again obtain 
\begin{equation}
     \int\limits_{-\infty}^{+\infty}dk_z\operatorname{Im}\Sigma^R_{\rm osc}(\Omega,k_z)\approx  - \frac{\pi k_F \bar{g}^2}{3}\frac{\nu_{\rm osc} }{\nu }|\Omega|,
\end{equation}
for $|\Omega|\ll \omega_c$. As before, the real part of the self-energy can be recovered by means of the Kramers-Kronig relation. In particular, for the odd part we obtain
\begin{equation}\label{eq:Omega_2/3}
\begin{aligned}
&[\operatorname{Re}\Sigma_{\rm osc}^R(\Omega,0)]_{\rm odd} = -\left(\frac{2\bar{g}^{4}\beta\omega_D}{\pi}\right)^{1/3}  \frac{\nu_{\rm osc}}{\nu}\operatorname{sgn}(\Omega) |\Omega|^{2/3},\\
&\int\limits_{-\infty}^{+\infty}dk_z\operatorname{Re}[\Sigma_{\rm osc}^R(\Omega,k_z)]_{\rm odd} = -\frac{2\bar{g}^2k_F}{3} \frac{\nu_{\rm osc}}{\nu} \Omega \ln \frac{\omega_c}{|\Omega|}, 
\end{aligned}
\end{equation}

where in the first line we used the integral identity \begin{equation}
    \frac{1}{\pi}\int\limits_{-\Lambda}^{\Lambda}\frac{d \omega |\omega|^{2/3}}{\omega-\Omega}= \sqrt{3} \operatorname{sgn}(\Omega)|\Omega|^{2/3}+\mathcal{O}\left(\frac{\Omega}{\Lambda}\right).
\end{equation}
We also note that the second line in Eq.~\eqref{eq:Omega_2/3} is exactly the same as Eq.~\eqref{eq:Sigma_k_z_int}, again up to an additional overall factor $1/3$.

There is also a way to obtain the same results by working on the Matsubara axis. Specifically, the oscillatory part of the Matsubara self-energy at $k_z=0$ reads as 
\begin{equation}\label{eq:Sigma_z=3_osc}
\begin{aligned}
&\Sigma_{\rm osc}(\ep,0) \approx \frac{2^{5/6}\bar{g}^{4/3}(\beta \omega_D)^{1/3}}{ 3\sqrt{3}\pi^{1/3} } \sqrt{\frac{\omega_c}{\mu}}\sum\limits_k \frac{(-1)^k}{k^{1/2}}\\
    & \times \int \frac{d\bar{\ep}\operatorname{sgn}(\bar{\ep})}{|\ep-\ep'|^{1/3}} \exp\Big\{-\frac{2\pi k }{\omega_c}|\bar{\ep}|\left(1+\frac{\bar{g}^2}{3}\ln\frac{\Lambda_D}{|\bar{\ep}|}\right)\\&+i\left(\frac{2\pi  k \mu}{\omega_c}-\frac{\pi}{4} \right)\operatorname{sgn}(\bar{\ep})\Big\}.
    \end{aligned}
\end{equation}
After expanding this expression in powers of $|\ep|\ll \omega_c$, we obtain
\begin{equation}\label{eq:Sigma_Mats_z=3}
    \Sigma_{\rm osc}(\ep,0) \approx  \Sigma_{\rm osc}(0,0) +\frac{2^{4/3}(\bar{g}^4\beta \omega_D)^{1/3}}{ \sqrt{3}\pi^{1/3} } \frac{\nu_{\rm osc}}{\nu} \operatorname{sgn}(\ep)|\ep|^{2/3}.
\end{equation}
Upon performing analytic continuation as $\Sigma^R(\Omega,k_z)= -i \Sigma(-i\Omega+0^+ ,k_z)$, we find an additional factor $\sqrt{3}/2$ ($1/2$) for the real (imaginary) part of $\Sigma^R(\Omega,k_z=0)$. The final result is thus in agreement with Eq.~\eqref{eq:Omega_2/3} and Eq.~\eqref{eq:Im_Sigma_k0z3}. A constant contribution $\Sigma_{\rm osc}(0,0)$ can be evaluated exactly at the leading order in $\bar{g}^2$. In this case, after setting $\ep=0$ in Eq.~\eqref{eq:Sigma_z=3_osc}, we obtain
\begin{equation}\label{eq:ReSigma00}
\begin{aligned}
    \operatorname{Re}\Sigma^R_{\rm osc}(0,0) &\approx \frac{  2^{7/6} \Gamma(2/3)}{3\sqrt{3}\pi } (\bar{g}^{4}\beta \omega_D \omega_c^2)^{1/3} \sqrt{\frac{\omega_c}{\mu}}
\\ &\times \sum\limits_{k}\frac{(-1)^k}{k^{7/6}}\sin\left(\frac{2\pi k \mu}{\omega_c}-\frac{\pi}{4}\right).\\
    \end{aligned}
\end{equation}
The asymptotic estimation of $\operatorname{Re}\Sigma^R_{\rm osc}(0,0)$ in the presence of the marginal FL self-energy in Eq.~\eqref{eq:Sigma_z=3_osc} (i.e. for $\bar{g}^2$ of the order of one) could be easily performed in the same fashion as it was done in Eq.~\eqref{eq:A_kT=0}. However, we emphasize that this would only reduce the value of $\operatorname{Re}\Sigma^R_{\rm osc}(0,0)$. As we discuss below, even in its present form, Eq.~\eqref{eq:ReSigma00} contributes only at the sub-leading order in $\omega_c/\mu$ to the final expression for the entropy, so there is no need to keep higher orders of $\bar{g}^2$.

It is clear from our derivation that the obtained low-frequency asymptotic form of $\Sigma^R(\Omega)$ does not change upon further iterations of the self-consistency equation even if the oscillations of $\Sigma^R(\Omega)$ itself are taken into account. Moreover, one could also check that Eq.~\eqref{eq:Sigma_Mats_z=3} survives the inclusion of the oscillatory part of the {\it{bosonic}} self-energy in Eq.~\eqref{eq:Sigma_eq_new}. The latter can be found by generalizing the derivation in \cite{Aleiner1995}. In the appropriate kinematic regime, we obtain (see Appendix \ref{sec:LD_B} for details)
\begin{equation}
    \begin{aligned}
        \Pi(\omega_m,q_\parallel, q_z)&\approx-\frac{\pi g^2\nu|\omega_m|}{2 v_F\sqrt{q_\parallel^2+q_z^2}}\left(1-\frac{(eB)^2}{4\pi k_Fq_\parallel^3}\right)\\
        &+\frac{g^2\nu|\omega_m|}{v_Fq_\parallel}\sqrt{\frac{\pi}{2}}\frac{\sin\left( 2R_cq_\parallel\right)}{\sinh(\pi\omega_m/\omega_c)}.
    \end{aligned}
\end{equation}
for $R_c^{-1}\ll q_\parallel,q_z\ll k_F$, and $|\omega_m|\ll v_F q_\parallel$. Indeed, the second term of the first line here is just a small anisotropic correction to the conventional Landau damping, whereas the second line has the same scaling, but quickly oscillates ($R_cq_\parallel \gg 1$), and thus its contribution to the integral over $q_\parallel$ is suppressed by additional powers of $\omega_c/\mu$.

\subsection{Oscillations of the entropy and magnetization}

Our next step to evaluate the fermionic contribution to the entropy in Eq.~\eqref{eq:entropy_large}. In contrast to the $z=1$ case, the oscillatory corrections to the self-energy turned out to be all small by an extra factor $\sqrt{\omega_c/\mu}$ compared to the smooth marginal FL self-energy, $\tilde{\Sigma}^R$, so we can simply expand Eq.~\eqref{eq:entropy_large} up to the lowest order in $\Sigma^R_{\rm osc}$. As a result, we find
\begin{widetext}
\begin{equation} \label{eq:entropy_large_new1}
\begin{aligned}
    &S_{\rm osc}^{(\rm el)} = \frac{(m\omega_c)^{3/2}}{4\pi^3} \sum\limits_{k=1}^{+\infty} \frac{(-1)^k}{k^{3/2}}   \int\limits_{-\infty}^{+\infty}d\Omega \frac{\partial n_F}{\partial T} e^{\frac{2\pi k}{\omega_c} \operatorname{Im}\tilde{\Sigma}^R(\Omega)}  \sin\left(\frac{2\pi  k} {\omega_c}\left[\Omega+\mu-\operatorname{Re}\tilde{\Sigma}^R(\Omega)\right]-\frac{\pi}{4}\right)\left\{1-\frac{2\pi k}{\omega_c} \operatorname{Im}\tilde{\Sigma}^R(\Omega)\right.\\
    & \left. -\frac{4\pi^2 k^2}{\omega_c^2} \operatorname{Im}\tilde{\Sigma}^R(\Omega)\operatorname{Im}\tilde{\Sigma}^R_{\rm osc}(\Omega,0)\right\} -\frac{m}{4\pi^2} \int\limits_{-\infty}^{+\infty}d\Omega \frac{\partial n_F}{\partial T}\int\limits_{-\infty}^{+\infty}dk_z \operatorname{Re}\Sigma^R_{\rm osc}(\Omega,k_z)\\
    &-\frac{m^{3/2}\omega_c^{1/2}}{2\pi^2} \sum\limits_{k=1}^{+\infty} \frac{(-1)^k}{k^{1/2}}   \hspace{-0.3em}\int\limits_{-\infty}^{+\infty}\hspace{-0.3em}d\Omega \frac{\partial n_F}{\partial T} e^{\frac{2\pi k}{\omega_c} \operatorname{Im}\tilde{\Sigma}^R(\Omega)}  \cos\left(\frac{2\pi  k} {\omega_c}\left[\Omega+\mu-\operatorname{Re}\tilde{\Sigma}^R(\Omega)\right]-\frac{\pi}{4}\right)\left(1-\frac{2\pi k}{\omega_c} \operatorname{Im}\tilde{\Sigma}^R(\Omega)\right)\operatorname{Re}\tilde{\Sigma}^R_{\rm osc}(\Omega,0)
  \end{aligned}
\end{equation}
As before, $\tilde{\Sigma}^R$ and $\tilde{\Pi}^R$ denote the `smooth' parts of the self-energy. First, we argue that all the term in Eq.~\eqref{eq:entropy_large_new1}  proportional to $\operatorname{Im}\tilde{\Sigma}^R(\Omega)=-\pi \bar{g}^2|\Omega|/6$ contribute only to a sub-leading temperature dependence which is even of higher order in terms of $T$-scaling than $\mathcal{O}(T)$. Indeed, we note that for each of this terms, after performing rescaling $\Omega=2Tx$, we obtain schematically $\sim T\int_{-\infty}^{+\infty}\frac{dx \; x|x|}{\cosh^2x} g(2T x)$ (note that this expression is already proportional to $T$). Here $g(\Omega)$ denotes the remaining sum over harmonics $k$. Thus, by parity, only the odd part of $g(2T x)$ contributes, which further increases the overall order in $T$. Therefore, we can neglect all such terms. This leaves us with the only three remaining terms
\begin{equation} \label{eq:entropy_large_new}
\begin{aligned}
    S_{\rm osc}^{(\rm el)} &\approx \frac{(m\omega_c)^{3/2}}{4\pi^3} \sum\limits_{k=1}^{+\infty} \frac{(-1)^k}{k^{3/2}}   \int\limits_{-\infty}^{+\infty}d\Omega \frac{\partial n_F}{\partial T} e^{\frac{2\pi k}{\omega_c} \operatorname{Im}\tilde{\Sigma}^R(\Omega)}  \sin\left(\frac{2\pi  k} {\omega_c}\left[\Omega+\mu-\operatorname{Re}\tilde{\Sigma}^R(\Omega)\right]-\frac{\pi}{4}\right)\\
    &-\frac{m^{3/2}\omega_c^{1/2}}{2\pi^2} \sum\limits_{k=1}^{+\infty} \frac{(-1)^k}{k^{1/2}}   \hspace{-0.3em}\int\limits_{-\infty}^{+\infty}\hspace{-0.3em}d\Omega \frac{\partial n_F}{\partial T} e^{\frac{2\pi k}{\omega_c} \operatorname{Im}\tilde{\Sigma}^R(\Omega)}  \cos\left(\frac{2\pi  k} {\omega_c}\left[\Omega+\mu-\operatorname{Re}\tilde{\Sigma}^R(\Omega)\right]-\frac{\pi}{4}\right)\operatorname{Re}\tilde{\Sigma}^R_{\rm osc}(\Omega,0)\\
    &-\frac{m}{4\pi^2} \int\limits_{-\infty}^{+\infty}d\Omega \frac{\partial n_F}{\partial T}\int\limits_{-\infty}^{+\infty}dk_z \operatorname{Re}\Sigma^R_{\rm osc}(\Omega,k_z).
  \end{aligned}
\end{equation}
\end{widetext}
The first term in Eq.~\eqref{eq:entropy_large_new}, which we denote as $S^{(\rm el,1)}_{\rm osc}$, essentially corresponds to the extended LK formula (cf. Eq.~\eqref{ampextendedlk}) but with a temperature derivative acting only on the distribution function $n_F(\Omega)$. Its low-temperature behavior can be easily computed in the same way as it was done in Sec.~\ref{Sec:Luttinger_expansion}, see Eq.~\eqref{eq:Extended_LK_A}. Here we just state the final result
\begin{equation}\label{eq:S_corr_N=1_1}
   S^{(\rm el,1)}_{\rm osc}\approx \frac{\pi^2 \bar{g}^2}{9} \nu_{\rm osc} T \ln \frac{\Lambda_D}{T}.
\end{equation} 
The remaining two terms in Eq.~\eqref{eq:entropy_large_new}, together denoted as $S^{(\rm el,2)}_{\rm osc}$, are analyzed in a similar fashion by performing a rescaling $\Omega=2Tx$ and expanding the integrand in the limit $T\rightarrow 0$. As a result, we obtain the following integral over $x$
\begin{equation}\label{eq:S_2_z=33}
    \begin{aligned}
        &S^{(\rm el,2)}_{\rm osc}\approx  -\nu\hspace{-0.3em}\int\limits_{-\infty}^{+\infty}\hspace{-0.5em}\frac{dx \; x}{\cosh^2 x}\left[\frac{2\bar{g}^2 xT}{3\nu} \frac{d\nu_{\rm osc}}{ d\mu} \ln\frac{\Lambda_D}{ 2T|x|} \operatorname{Re}\Sigma^{R}_{\rm osc}(0,0) 
         \right.\\
        & \left.+ \frac{ \nu_{\rm osc}}{\nu}\operatorname{Re}\Sigma^{R}_{\rm osc}(2Tx,0)+\int\limits_{-\infty}^{+\infty} \frac{dk_z}{2k_F} \operatorname{Re}\Sigma^{R}_{\rm osc}(2T x,k_z)\right],
    \end{aligned}
\end{equation}
where the derivative $d/d\mu$ in the last line is acting only on $\mu$ in the argument of $\cos(...)$ in Eq.~\eqref{eq:compress22}. Next, one can easily see that even though the first term in Eq.~\eqref{eq:S_2_z=33} in combination with Eq.~\eqref{eq:ReSigma00} scales as $T\ln T$, it has an overall prefactor $(\omega_c/
\mu)^{2/3}$, as opposed to the last term in Eq.~\eqref{eq:S_2_z=33} which also scales as $T\ln T$ but is of the order of $(\omega_c/\mu)^{1/2}$. Thus, we ignore this higher-order correction and obtain
\begin{equation}\label{eq:S_2_z=3}
\begin{aligned}
    S^{(\rm el,2)}_{\rm osc}\approx \frac{\pi^2\bar{g}^2}{9} \nu_{\rm osc} T\ln \frac{\omega_c}{T} +\frac{4c_2\bar{g}^{4/3}}{\pi^{1/3}}(\beta \omega_D)^{1/3} \frac{\nu_{\rm osc}^2}{\nu} T^{2/3}
    \end{aligned}
\end{equation}
where $c_2=\int_0^{+\infty} dx x^{5/3}/\cosh^2x \approx 0.74$. After combining Eqs.~(\ref{eq:S_corr_N=1_1},\ref{eq:S_2_z=3},\ref{eq:S_4_z=3}), we finally arrive at the following result for the total entropy
\begin{equation}\label{eq:S_DHVA_z=3}
\begin{aligned}
     S^{}_{\rm osc}&=S^{\rm(bos)}_{\rm osc}+S^{\rm(el,1)}_{\rm osc}+S^{\rm(el,2)}_{\rm osc} \\ &\approx
    \frac{2\pi^2\bar{g}^2}{9} \nu_{\rm osc} T\ln \frac{\sqrt{\omega_c \Lambda_D}}{T} \\
    &+\frac{4c_2\bar{g}^{4/3}}{\pi^{1/3}}(\beta \omega_D)^{1/3} \frac{\nu_{\rm osc}^2}{\nu} T^{2/3}\\
     &+\frac{8c_1\pi^{1/3}\bar{g}^{8/3}}{9\sqrt{3}}  (\beta \omega_D)^{2/3}\nu_{\rm osc}  T^{1/3},
\end{aligned}
\end{equation}
and the total entropy is given by the sum of the smooth and oscillatory contributions $S=\frac{\pi^2 \bar{g}^2}{9} \nu T \ln (\Lambda_D/T) + S^{}_{\rm osc}$. It is worth noting that the last term $\sim T^{1/3}$ in Eq.~\eqref{eq:S_DHVA_z=3} can exceed the smooth contribution $\sim T \ln (\Lambda_D/T)$ at sufficiently low temperatures. This, in turn, could indicate a potential thermodynamic instability towards an ordered state at certain values of $\mu/\omega_c$ such that $\nu_{\rm osc}$ is negative. In this case, the total compressibility in a field $\nu+\nu_{\rm osc}$ is less than the bare compressibility $\nu$, which we used to tune to the critical point at $B=0$. As a consequence, the boson condenses at low temperatures since the actual position of the critical point is shifted by the magnetic field.

The DHVA magnetization is obtained via a Maxwell relation as
\begin{equation}\label{eq:DHVA_z=3}
\begin{aligned}
     M^{}_{\rm osc}&\approx
    M_{\rm osc}(T=0)+\frac{\pi^2\bar{g}^2}{9} \frac{d\nu_{\rm osc}}{dB} \;T^2\ln \frac{\sqrt{\omega_c \Lambda_D}}{T} \\
    &+\frac{24c_2\bar{g}^{4/3}}{5\pi^{1/3}}(\beta \omega_D)^{1/3} \frac{\nu_{\rm osc}}{\nu} \frac{d\nu_{\rm osc}}{dB}\;T^{5/3}\\
&+\frac{2c_1\pi^{1/3}\bar{g}^{8/3}}{3\sqrt{3}}  (\beta \omega_D)^{2/3}\frac{d\nu_{\rm osc}}{dB} \; T^{4/3}.
\end{aligned}
\end{equation}
The unconventional low-temperature power-law asymptotic $\sim T^{5/3}$ and  $\sim T^{4/3}$ of the DHVA oscillations is the main result of this section. At the lowest temperatures, the $\sim T^{4/3}$ term dominates, and the thermodynamic potential asymptotically behaves as in  Eq.~\eqref{eq:Xi_0_osc_form}, where the amplitude $A_k(T)$ has the form
\be\label{eq:A_z=3_final}
A_k(T)-A_k(T=0)\approx -\frac{2c_1\pi^{1/3}\bar{g}^{8/3} k}{3\sqrt{3} \omega_c} (\beta\omega_D)^{2/3}T^{4/3}.
\ee
We will discuss Eq.~\eqref{eq:S_DHVA_z=3} and Eq.~\eqref{eq:DHVA_z=3} in the next section, and compare them to the $z=1$ result Eq.~\eqref{eq:DHVA_z=1} obtained in the previous section.

\section{Discussion and conclusions}\label{sec:Discussion}

In this work, we have developed a new theory of the DHVA oscillations that can capture both conventional Fermi liquid regime and, more crucially, the quantum critical regime for which the Lifshitz-Kosevich paradigm fails to work due to singular fermion interactions mediated by massless critical bosons. The naive adoption of the LK paradigm at criticality results in the violation of the third law of thermodynamics.  We have shown that at a quantum critical point,  the breakdown of Luttinger's extension to LK's theory is due to the oversimplification of neglecting the oscillatory part of the fermion self-energy, and the bosonic contribution.  
By contrast, our approach, which is based on the analysis of the Luttinger-Ward functional within the Migdal-Eliashberg approximation, is  capable of treating the fermionic and bosonic contributions on equal footing and manifestly satisfies the third law of thermodynamics.


The oscillatory magnetization resulting from a thermodynamically viable entropy exhibits significant deviations from the standard LK theory, especially at low temperatures $T \ll \omega_c$, and the leading temperature dependence of the oscillation amplitude $A_k(T)$ depends on the dynamical scaling of the critical bosonic mode. In case of the undamped $z=1$ boson, we find that $A_k(T)$ exhibits a $T^{2}\ln(\omega_c/T)$ attenuation which differs from the usual LK behavior $T^2$ by an extra logarithmic factor $\sim \ln(\omega_c/T)$ (see Eq.~\eqref{eq:A_z=1_final}). The origin of this non-analytic contribution can be traced back to the ‘oscillatory’ part of the fermionic self-energy averaged over all values of $k_z$, which retains a marginal Fermi liquid form even at frequencies smaller than $\omega_c$, see Eq.~\eqref{eq:ImSigmakzaveragedz=1}. We also note the proposed $\sim T^{2}\ln(\omega_c/T)$ low-temperature asymptotic behavior of $A_1(T)$ formally resembles the results of \cite{Pelzer1991,Wasserman_1991} obtained from a different phenomenological model.

In contrast, the Landau overdamped $z=3$ critical boson leads to several non-analytic low-$T$ contributions, with the most dominant one scaling as $T^{4/3}$ at lowest temperatures, see Eq.~\eqref{eq:A_z=3_final}. The latter contribution is formally associated with the bosonic part of the thermodynamic potential, and results from the combination of the $z=3$ dynamical scaling and the finite oscillatory part of the compressibility $\nu_{\rm osc}$, cf. Eq.~\eqref{eq:compress22}. In addition, Eq.~\eqref{eq:DHVA_z=3} also contains a sub-leading term $\sim T^{5/3}$ stemming from the oscillatory part of the fermionic self-energy at the extremal orbit. Its non-analytic behavior is related to the `dimensional reduction' of the phase volume available for scattering processes that do not take the fermion away from the extremal orbit, see Fig.~\ref{fig:dim_red} and discussion in Sec.~\ref{sec:dim_red_sec}. It is also worth mentioning that this contribution appears with an unconventional oscillating pre-factor $\nu_{\rm osc}\frac{d\nu_{\rm osc}}{dB}$ (all other terms in our Eq.~\eqref{eq:DHVA_z=3} or in the extended LK theory are accompanied by a single factor of $\nu_{\rm osc}$ only). The appearance of {\it{two}} factors of $\nu_{\rm osc}$ can already be seen perturbatively, by considering the lowest-order diagram for the LW functional $\Phi$ and singling out the oscillatory contribution from the Poisson summations of {\it both} Green's functions at the same time. In the conventional Luttinger's expansion, such contributions are of the order of $\sim T^2$ and further suppressed by an extra factor of $\omega_c/\mu$. However, here they produce a strong non-analytic temperature dependence, and thus, must be retained. It is also worth mentioning that if the system is slightly detuned from a quantum critical point (i.e. if the dressed bosonic propagator has a small but finite physical `mass' $M_{ b}=\sqrt{m_b^2-g^2\nu}>0$), then at sufficiently low temperatures the conventional LK-type behavior is recovered $A_k(T)-A_k(T=0)\sim -\bar{g}^2 k \omega_D^2 T^2/(M_{ b}^2\omega_c)$. As a result, the anomalous temperature dependence of the oscillation amplitudes in Eq.~\eqref{eq:A_z=3_final} should be cut off at the crossover scale that could be estimated as $T_{\rm cross}\sim \bar{g}\beta M_b^3/\omega_D^2$.


For higher temperatures $T\gtrsim \omega_c$ (but still $T\ll \omega_D,\mu$), we find that the tail of the oscillation amplitude $A_k(T)$ is strongly smeared by thermal fluctuations in both $z=1$ and $z=3$ cases and has the asymptotic form $T\exp\{-\# (T/\omega_c) \ln\omega_D/T\}$, cf. Eq.~\eqref{eq:A_high_T}, as opposed to the conventional LK prediction $T\exp\{-\# T/\omega_c\}$. We note that this result could be interpreted as an effective temperature-dependent Dingle factor resulting from scattering by `static' fluctuations of the order parameter. Our main findings in both regimes $T\ll \omega_c$ and $T\gtrsim \omega_c$ are summarized in Fig.~\ref{fig:intro}. Our theory can be applied to three-dimensional metals near quantum critical points, which exhibit marginal Fermi liquid behavior, such as ZrZn$_2$, an itinerant electron ferromagnet\footnote{We thank Erez Berg for bringing this system to our attention}.  A search for deviations of the LK behavior in this system would be an interesting direction for future experimental investigation. In order to make a more detailed comparison with experiments, one also would have to account for possible deviations from strict parabolicity of the bare electron dispersion, in the same way as it was done in the conventional LK theory. In principle, this extension could be analyzed in full detail based on Eqs.~(14),(17), and (49) which remain valid for arbitrary dispersion. However, we expect the temperature dependence of the DHVA amplitude at low temperatures $T\ll \omega_c\ll \omega_D$ to remain the same since it is supposed to be sensitive only to the universal dynamical scaling laws and phase volume arguments.

A natural extension of our theory would involve the study of quantum oscillations in two-dimensional (2D) metals, where non-Fermi liquid behavior is a far more singular phenomenon. It is possible that magneto-oscillations in 2D exhibit more conventional behavior at low temperatures due to the fact that the Landau level spectrum is gapped in 2D.  By contrast, in three dimensions, gapless excitations persist in the third direction, along the magnetic field, and produce singular effects, as we have shown here.  However, it is known that there are some subtleties even in the case of two-dimensional Fermi liquids, as discussed in Refs. 
\cite{Miyake1993,Stamp1998, Maslov2003}. In particular, one should distinguish between the situation when the chemical potential is fixed between Landau levels, and when the total number of electrons is fixed instead (i.e. there is a partially-occupied Landau level) \cite{Miyake1993}. In the latter case, it is known \cite{Aleiner1995} that the electron dynamics at energies less than $\omega_c$ (which ultimately determines thermodynamic properties at $T\ll \omega_c$) is rather intricate even for a weakly interacting Fermi gas. At the same time, the thermal smearing of quantum oscillations at $T\gtrsim \omega_c$ is expected to be much stronger in 2D than in 3D due to the singular static contributions to the self-energy at the first Matsubara frequency \cite{PhysRevB.102.045147, KleinFiniteT_SE}. Another intriguing future direction would be to study the effect of quenched disorder on DHVA oscillations in our three-dimensional marginal Fermi liquid model. Beyond the appearance of the conventional Dingle factor \cite{shoenberg_1984,abrikosov2017fundamentals}, one could expect some additional temperature-dependent renormalization effects of the
effective fermion mass due to the interplay of the critical interaction and impurity scattering \cite{Maslov2003,Adamov2006}. Finally, it would be interesting to investigate how quantum critical fluctuations affect the emergence of the diamagnetic Condon domains \cite{Condon} when the correlation length exceeds the cyclotron radius \footnote{We thank Leonid Levitov for bringing this problem to our attention}.



\begin{acknowledgements}
We thank E.~Berg, I.S.~Burmistrov, A.~Chubukov, S.~Kivelson, V.~Kravtsov, and C.M.~Varma for useful discussions. P.A.N. acknowledges the hospitality extended to him during his time as a Graduate Fellow at the Kavli Institute for Theoretical Physics, where the final stages of this work were completed. This research was supported in part by the National Science Foundation under Grant No. NSF PHY-1748958 and NSF PHY-2309135, the Heising-Simons Foundation, and the Simons Foundation (216179, LB). The work of P.A.N. and S.R. was supported in part by the US Department of Energy, Office of Basic Energy Sciences, Division of Materials Sciences and Engineering, under contract number DE-AC02-76SF00515. YMW acknowledges the Gordon and Betty Moore Foundation’s
EPiQS Initiative through GBMF8686 for support. 
\end{acknowledgements}

\appendix
\section{Oscillatory part of the self-energy for electron-phonon interaction}\label{sec:phonons}
In this section, we replace the critical interaction in Eq.~\eqref{eq:Sigma_eq_new} with a simple optical phonon propagator
\begin{equation}
    D_0(\omega_m,q) = \frac{1}{\omega_m^2+\omega_D^2},
\end{equation}
and analyze the oscillatory part of the fermionic self-energy arising in the limit $\mathcal{N}\gg 1$, and to all orders in the coupling strength $g^2$. The corresponding contribution to the thermodynamic potential at the lowest order in $g^2$ was first discussed in \cite{Engelsberg1970}. At $B=0$, the self-consistency equation is essentially one-loop exact, and the Matsubara self-energy reads as
\begin{equation}\label{eq:Sigma_phonons_B=0}
\begin{aligned}
    &\Sigma(\ep_n)=\pi\nu g^2T \sum\limits_{m} \frac{\operatorname{sgn}(\ep_m)}{(\ep_n-\ep_m)^2+\omega_D^2}\\
    &= \frac{\nu g^2}{2\omega_D} \operatorname{Im}\left\{\psi\left(\frac{i\omega_D-\ep_n}{2\pi T}+\frac{1}{2}\right)-\psi\left(\frac{i\omega_D+\ep_n}{2\pi T}+\frac{1}{2}\right)\right\}
    \end{aligned}
\end{equation}
Here $\nu=k_F^2/(2\pi^2v_F)$ is the density of states at the Fermi level. In particular, for the first Matsubara frequency, we find
\begin{gather}
      \Sigma(\pi T)=\pi \nu g^2 T/\omega_D^2.
\end{gather}
Here we made use of some properties of the digamma function
\begin{equation}
\begin{aligned}
 &\operatorname{Im}\psi(\sigma+ib) = \frac{\pi}{2}\coth\left(\pi b\right)+\frac{(-1)^\sigma}{2b},\quad \sigma=0,1\\
&\operatorname{Im}\psi\left(\frac{1}{2}+ib\right) =\frac{\pi}{2}\tanh(\pi b),
\end{aligned}
\end{equation}
for $b\in R$. The absence of any momentum dependence of the bare propagator dramatically simplifies the self-consistency equation in the presence of the field Eq.~\eqref{eq:Sigma_eq_new}. Indeed, the integral over $q_\parallel$ just gives the normalization condition for the associated Laguerre polynomials $l_B^2\int_0^{+\infty} dq_\parallel \;q_\parallel X_{nn'}(q_\parallel)=1$ independent of $n$ and $n'$. After making use of the Poisson summation formula Eq.~\eqref{eq:Poisson}, and performing the $q_z$ integral by the stationary phase method, we obtain
\begin{equation}\label{eq:Sigma_phonons_full}
\begin{aligned}
    \Sigma(\ep_m)
    \approx \Sigma(\ep_m)_{B=0} +
     \sqrt{\frac{\omega_c}{\mu}} \sum\limits_{k=1}^{+\infty}\frac{(-1)^k}{\sqrt{2k}} \sigma_k(\ep_m),
    \end{aligned}
\end{equation}
where the self-consistency condition for the amplitudes is given by
\begin{equation}\label{eq:SCBA_optical_phonons}
\begin{aligned}
    &\sigma_k(\ep_m)=\pi\nu g^2 T\sum\limits_{\bar{m}}  \frac{\operatorname{sgn}(\ep_{\bar{m}})}{(\ep_m-\ep_{\bar{m}})^2+\omega_D^2}
 \\
 &\times \exp\left\{-\frac{2\pi k}{\omega_c}|\ep_{\bar{m}}+\Sigma(\ep_{\bar{m}})|+i\left(\frac{2\pi  k \mu}{\omega_c}-\frac{\pi}{4}\right) \operatorname{sgn}(\ep_{\bar{m}})\right\} 
    \end{aligned}
\end{equation}
In principle, Eq.~\eqref{eq:Sigma_phonons_full} and \eqref{eq:SCBA_optical_phonons} result into an infinite set of equations for the amplitudes $\sigma_k(\ep_m)$, $k=1,2,...$. However, if we are only interested in the leading order correction in terms of $\omega_c/\mu \ll 1$, then it is sufficient to replace the full $\Sigma(\ep_{\bar{m}})$ in Eq.~\eqref{eq:SCBA_optical_phonons} by its zero field value $\Sigma(\ep_{\bar{m}})_{B=0}$ computed in Eq.~\eqref{eq:Sigma_phonons_B=0}. In this case, it is easy to see that $\sigma_k(\ep_m)$ is of the order of $\omega_c$ due to the exponential suppression factor. Thus, the full oscillatory part of the self-energy (the second term in Eq.~\eqref{eq:Sigma_phonons_full}) is of the order of $(\omega_c/\mu)^{3/2}$ compared to $\Sigma(\ep_{m})_{B=0}$.

\section{Self-energy in the absence of the field}
\label{sec:self-energy}
In this section, we revisit the calculation of the fermionic finite-temperature self-energy at criticality in the absence of the magnetic field, paying special attention to the thermal part of the self-energy.
We first show the calculations for the Landau-overdamped boson case which leads to dynamical scaling exponnet $z=3$, and then discuss undamped boson case where $z=1$. In both situations, the fermion self energies behave like marginal Fermi liquid. 

\subsection{Landau-overdamped limit: $z=3$}
The full fermionic self-energy in Eq.\eqref{selfconsistentzerofield} acquires the following form in the momentum space
\begin{equation}\label{eq:SCBA}
\begin{aligned}
    \Sigma(\ep_n)&=ig^2T \sum\limits_{m} \int\frac{d^3q}{(2\pi)^3} \frac{1}{(m_b^2+c^2q^2+\omega_m^2-\Pi(\omega_m,q))} \\
    &\times \frac{1}{i\ep_n+i \omega_m+i \Sigma(\ep_n+\omega_m)  -\xi_{k+q}}\;,
\end{aligned}
\end{equation}
where $\xi_k=k^2/2m-\mu$. In principle, the self-energy also has a very weak (and regular) momentum dependence near the Fermi surface, so we ignore it.  We also assume that the boson velocity is much smaller than the Fermi velocity, $c\ll v_F$. In this case, defining $\beta=v_F/c \gg 1$, the effective ``Debye frequency", $\omega_D=E_F/\beta \ll E_F$, i.e. the transferred energy is much smaller than the Fermi energy. In addition, Eq.~\eqref{eq:SCBA} should be supplemented by the equation for the bosonic self-energy $\Pi(\omega_m,q)$, 
\begin{equation}\label{eq:Pi_Sigma}
\begin{aligned}
    &\Pi(\omega_n,q) = -g^2 T \sum\limits_m \int\frac{d^3k}{(2\pi)^3}\frac{1}{(i\ep_m+i\Sigma(\ep_m) -\xi_{k})}\\
    &\times \frac{1}{i\omega_n+i\ep_m+i\Sigma(\omega_n+\ep_m) -\xi_{k+q}}=g^2\nu-\\
    & - \frac{2\pi g^2\nu T}{v_F q} \sum\limits_{m=0}^{|n|-1} \arctan \left(\frac{v_Fq}{\Sigma(\ep_m)+\Sigma(|\omega_n|-\ep_m)+|\omega_n|}\right),
    \end{aligned}
\end{equation}
assuming that $q\ll k_F$. In the absence of the self-energy effects (i.e. in the RPA-type approximation), we recover the usual expression for the polarization function of a three-dimensional electron gas at small momenta $q\ll k_F$
\begin{equation}\label{eq:apprndix_Damping}
\Pi(\omega_n,q)\approx g^2\nu\left(1-\frac{|\omega_n|}{v_Fq}\arctan \frac{v_Fq}{|\omega_n|}\right).
\end{equation}
where the second term leads to the usual Landau damping of the boson due to its decay into a particle-hole continuum. The condition for criticality is then given by $m_b^2-\Pi(0,q\rightarrow 0)=0$. In the regime $v_F q\gg |\omega_n|$, we find
\begin{equation}
    D(\omega_n,q)\approx \frac{1}{c^2q^2+\frac{\pi g^2\nu}{2v_F}\frac{|\omega_n|}{q}}\;.\label{eq:z3B5}
\end{equation}
From this expression, we read the usual $z=3$ dynamical scaling associated with Landau damping.

In principle, Eq.~\eqref{eq:SCBA} should be now solved together with Eq.~\eqref{eq:Pi_Sigma} as a coupled system of equations. However, it turns out \cite{Damia2019} that the fermionic self-energy computed with this RPA-dressed bosonic propagator is asymptotically self-consistent in the low-energy limit. Inserting Eq.\eqref{eq:z3B5} into Eq.~\eqref{eq:SCBA} and performing the momentum integration, we obtain
\begin{equation}
\begin{aligned}
    \Sigma(\ep_n)&=2\bar{g}^2T \operatorname{Im}\operatorname{Li}_2\left(\frac{i\beta \omega_D }{\ep_n+\Sigma(\ep_n)}\right)\\
    &+\frac{\pi\bar{g}^2T }{3}\sum\limits_{m\neq 0} \operatorname{sgn}\left(\ep_n+\Omega_m\right) \ln\left(1+ \frac{\Lambda_D}{|\omega_m|}\right)\;,
    \end{aligned}
\end{equation}
where $\Lambda_D= 2\beta \omega_D^3/(\pi g^2\nu)$ and $\bar{g}^2=g^2/(4\pi^2 c^2 v_F)$ is the dimensionless coupling constant.  $\operatorname{Li}_2(x)$ is the polylogarithm. Since $\Sigma(-\ep_n)=-\Sigma(\ep_n)$, we consider the case $n\geq 0$. Assuming that $\ep_n+\Sigma(\ep_n)\ll E_F$, we can approximate $\operatorname{Im}\operatorname{Li}_2(ix)\approx (\pi/2)\ln x$, and evaluate the sum explicitly. The result can be written in a compact way as
\begin{equation} \label{eq:Sigma_approx_eq_SCBA}
    \Sigma(\ep_n)=\pi\bar{g}^2 T\ln \left(\frac{\beta\omega_D \zeta_n}{\ep_n+\Sigma(\ep_n)}\right)\;,\quad n\geq 0\;,
\end{equation}
where we defined
\begin{equation}
    \zeta_n=\left(\frac{\Gamma(1+\frac{\Lambda_D}{2\pi T}+n)}{n!\Gamma(1+\frac{\Lambda_D}{2\pi T})}\right)^{2/3}.
\end{equation}
Here $\Gamma(x)$ is the Gamma function. The equation above admits a solution in terms of the Lambert function $W(x)$ (`product log')
    \begin{equation}
     \Sigma(\ep_n)=-\ep_n+\pi\bar{g}^2T W\left(\frac{\beta\omega_D}{\pi\bar{g}^2T} \zeta_n \exp\left\{\frac{\ep_n}{\pi\bar{g}^2T}\right\}\right).
    \end{equation}
For large $n$, we use the approximation 
\begin{equation}
    \zeta_n \approx  \frac{1}{(n!)^{2/3}} \left(\frac{\Lambda_D}{2\pi T}\right)^{2n/3}\;, \quad n\geq 1,\quad T\ll \Lambda_D.
\end{equation}
At the first Matsubara frequency, $\ep_0=\pi T$, and in the limit of low temperature $T$, we obtain
\begin{equation}
\label{eq:Sigma_first_Matsubara}
    \Sigma(\pi T) \approx \pi \bar{g}^2 T\ln \left(\frac{\beta \omega_D}{\pi\bar{g}^2 T}\right),
\end{equation}
where we used $ W(x)\approx \ln x-\ln\ln x+o(1)$ for $x\gg 1$. Note that this expression is non-analytic as a function of $\bar{g}^2$ and stems from the thermal mode alone since the dynamical contribution (in the second line of Eq.\eqref{eq:Sigma_eq_Li}) vanishes at $n=-1,0$ (this can be seen from the fact that $\zeta_0=1$), as expected from the first Matsubara frequency rule \cite{Chubukov_first_Matsubara}.

Next, let us consider the limit when $\ep_n$ is fixed, and $T$ goes to zero. In this case, we obtain 
\begin{equation}\label{eq:Sigma_z=3_B=0_full}
    \begin{aligned}
        &\Sigma(\ep_n) = \Sigma_0(\ep_n) +\pi \bar{g}^2 T\ln \left(\frac{(\pi\beta^2 \bar{g}^2 \omega_D^2 T)^{1/3} }{\ep_n+\pi\bar{g}^2T \ln \frac{\beta}{\pi\bar{g}^2} +\Sigma_0(\ep_n)}\right)\\
        &\Sigma_0(\ep_n)=\frac{\bar{g}^2}{3} \ep_n \ln \left(\frac{\Lambda_D}{2\pi T}\right) -\frac{2\pi\bar{g}^2}{3}T \ln \left[\frac{\Gamma\left(\frac{\ep_n}{2\pi T}+\frac{1}{2}\right)}{\sqrt{2\pi}}\right].
    \end{aligned}
\end{equation}
As a special case, if $T\ll \ep_n$, then we can expand
\begin{equation}\label{eq:B13}
    \Sigma_0(\ep_n) \approx \frac{1}{3}\bar{g}^2\ep_n\ln\frac{\Lambda_D}{\ep_n}+\frac{1}{3}\bar{g}^2 \ep_n +\mathcal{O}(T^2)\;.
\end{equation}
And thus, we obtain the following asymptotic expression 
\begin{equation}\label{eq:Sigma_T_approx_0}
     \Sigma(\ep_n) \approx \Sigma_0(\ep_n)_{T=0} +\pi \bar{g}^2 T\ln \left(\frac{(\pi\beta^2 \bar{g}^2 \omega_D^2 T)^{1/3}}{\ep_n +\Sigma_0(\ep_n)_{T=0}}\right),
\end{equation}
for $T\ll \ep_n$. 
The dynamical part of the self-energy, $\Sigma_0(\ep_n)$, leads to the usual marginal Fermi liquid behaviour.

\subsection{Undamped limit: $z=1$}
In this limit, there is no feedback from the fermions to the bosons, and the bosonic self-energy $\Pi(\omega_m,q)$ can be neglected. Thus, the integral equation \eqref{eq:SCBA} represents a sum over the ``rainbow'' diagrams. The condition for criticality is then simply $m_b^2=0$.

At zero temperature, the momentum integration in Eq.~\eqref{eq:SCBA} can be carried out by treating the density of states as a constant \cite{Schrieffer}. In this case, solution of Eq.~\eqref{eq:SCBA} turns out to be one-loop exact: higher order rainbow diagrams contain pairs of Green's functions with the same momentum and frequency, and thus the integration over their momenta vanish since both poles occur in the same complex half-plane. At finite temperature, the situation is more delicate due to the discrete nature of the Matsubara frequencies. Indeed, Eq.~\eqref{eq:SCBA} always contains a term with $m=0$ exhibiting an IR divergence $\sim 1/q^2$ from the static bosonic fluctuations. The standard way of dealing with this problem is to split the self-energy into the 'thermal' ($m=0$) and 'dynamical' ($m\neq0$) contributions, and treat the density of states as a constant only in the latter, whereas in the former the full momentum integration is performed \cite{Wang2017,KleinFiniteT_SE}. This procedure leads to the following equation
\begin{equation}
\begin{aligned}\label{eq:Sigma_eq_Li}
    \Sigma(\ep_n)&=2\bar{g}^2T \operatorname{Im}\operatorname{Li}_2\left(\frac{i\beta \omega_D }{\ep_n+\Sigma(\ep_n)}\right)\\
    &+\frac{\pi\bar{g}^2T }{2}\sum\limits_{m\neq 0} \operatorname{sgn}\left(\ep_n+\Omega_m\right) \ln\left(1+ \frac{\omega_D^2}{\omega_m^2}\right)\;\\
    &=\pi\bar{g}^2 T\ln \left(\frac{\beta\omega_D \zeta_n}{\ep_n+\Sigma(\ep_n)}\right)
    \end{aligned}
\end{equation}
which is similar to Eq. but with a different $\zeta_n$ given by
\begin{equation}
    \zeta_n=\frac{\sinh (\omega_D/2T)}{(\omega_D/2T)} \frac{|\Gamma(1+n+\frac{i\omega_D}{2\pi T})|^2}{\Gamma^2(n+1)}\;.
\end{equation}
For large $n$, we find
\begin{equation}
    \zeta_n \approx \frac{1}{(n!)^2} \left(\frac{\omega_D}{2\pi T}\right)^{2n}\;, \quad n\geq 0,\quad T\ll \omega_D\;.
\end{equation}   

The solution of $\Sigma(\varepsilon_n)$ is similar to $z=3$ case in the limit when $\ep_n$ is fixed, and $T$ goes to zero. The results are 
\begin{equation}\label{eq:Sigma_full_Ngg1}
    \begin{aligned}
        &\Sigma(\ep_n) = \Sigma_0(\ep_n) +\pi \bar{g}^2 T\ln \left(\frac{\beta T}{\ep_n+\pi\bar{g}^2T \ln \frac{\beta}{\pi\bar{g}^2} +\Sigma_0(\ep_n)}\right)\\
        &\Sigma_0(\ep_n)=\bar{g}^2 \ep_n \ln \left(\frac{\omega_D}{2\pi T}\right) -2\pi\bar{g}^2T \ln \left[\frac{\Gamma\left(\frac{\ep_n}{2\pi T}+\frac{1}{2}\right)}{\sqrt{2\pi}}\right].
    \end{aligned}
\end{equation}
This can be compared to Eq.\eqref{eq:Sigma_z=3_B=0_full}
As a special case, if $T\ll \ep_n$, then we can expand
\begin{equation}\label{eq:Sigma_0_small_T}
    \Sigma_0(\ep_n) \approx \bar{g}^2\ep_n\ln\frac{\omega_D}{\ep_n}+\bar{g}^2 \ep_n +\mathcal{O}(T^2)\;.
\end{equation}
Comparing with Eq.\eqref{eq:B13}, we see that both give rise to marginal Fermi liquid behavior, and the only differences are a factor of $1/3$ in the $z=3$ case, and the cutoffes $\Lambda_D$ versus $\omega_D$.
And thus, we obtain the following asymptotic expression for $z=1$ case:
\begin{equation}\label{eq:Sigma_T_approx_0}
     \Sigma(\ep_n) \approx \Sigma_0(\ep_n)_{T=0} +\pi \bar{g}^2 T\ln \left(\frac{\beta T}{\ep_n +\Sigma_0(\ep_n)_{T=0}}\right),
\end{equation}
for $T\ll \ep_n$. One can also directly verify this result by substituting it back into the r.h.s. of Eq.~\eqref{eq:Sigma_approx_eq_SCBA}, expanding in $T$, and making use of the following asymptotic expression
\begin{equation}
    \ln \zeta_n=\frac{\ep_n}{\pi T}+\frac{\ep_n}{\pi T}\ln \frac{\omega_D}{\ep_n}+\ln \frac{T}{\omega_D}+\mathcal{O}(T)\;.
\end{equation}
The dynamical part of the self-energy, $\Sigma_0(\ep_n)$, leads to the usual marginal Fermi liquid behavior.

\begin{figure}
    \includegraphics[width=8cm]{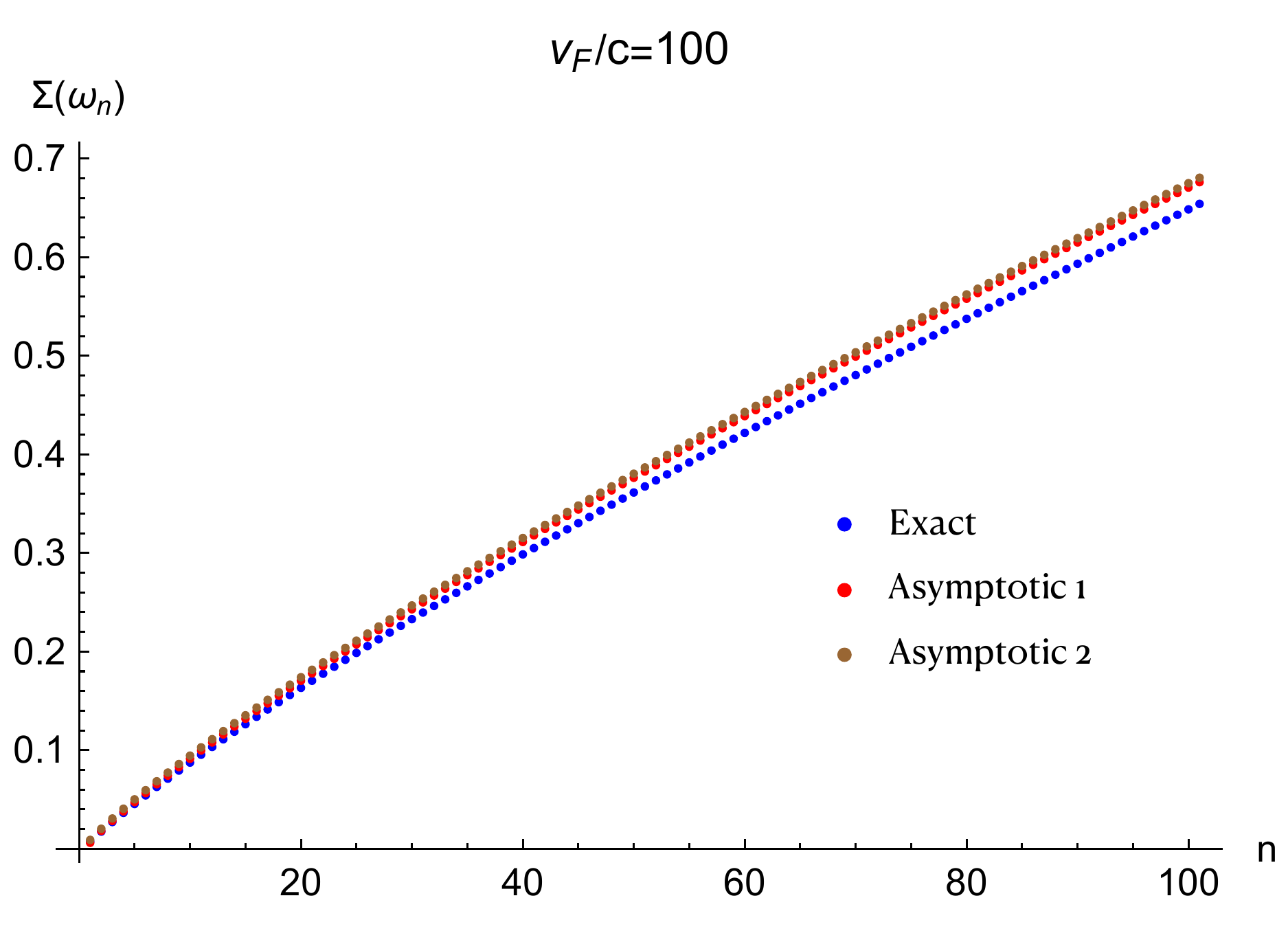} 
    \caption{Self energy obtained in different ways at $T=2\times10^{-4}$. All the energy scales are measured in units of $g^2$, and we have made the dimensionless coupling $\bar{g}^2=1$. Here the exact result (blue) is obtained by directly solving Eq.\eqref{eq:SCBA} in the limit of $m_b\to0$ and neglecting $\Pi$. The first aymptotic result (red) is obtained from numerically solving Eq.\eqref{eq:Sigma_eq_Li}, which does not require iteration. And the second asymptotic result (brown) is from Eq.\eqref{eq:Sigma_T_approx_0}. }
    \label{fig:compare}
\end{figure}

Complementary to the above analytical results, the solution to the self energy can also be obtained numerically, even without any approximation. As a test of the realiability of our analytical results, we show in Fig.\ref{fig:compare} a comparison between numerical results and Eq.\eqref{eq:Sigma_T_approx_0} for a small enough $T$. The exact solution to Eq.\eqref{eq:SCBA} in the limit of $m_b\to0$ and neglecting $\Pi$ is found by numerical iteration, for which the momentum integrations in both thermal and dynamical part is treated in the same way.  We also solve Eq.\eqref{eq:Sigma_eq_Li}, which is an approximated equation in which the momentum integration for the dynamical part is carried out by separating the tangential and perpendicular direction with respect to the local Fermi surface. This simplified equaiton does not require iteration since $\Sigma(\varepsilon_n)$ is determined by quantities only at $\varepsilon_n$. And lastly the analytical results in Eq.\eqref{eq:Sigma_T_approx_0} is plotted for comparison. We see a good agreement among these different approaches.

\section{Landau damping in a magnetic field}\label{sec:LD_B}
In this section, we revisit the calculation of the one-loop bosonic self-energy in Eq.~\eqref{eq:Pi_eq} in the presence of a weak magnetic field and in three dimensions. The analogous result in two dimensions was obtained in Ref.~\cite{Aleiner1995}, and here we follow their approach. We first recall that in a three-dimensional system without a magnetic field, the bosonic self-energy $\Pi(\omega_m,q)$ on the Matsubara axis is given by Eq.~\eqref{eq:apprndix_Damping}, where the small frequency and momentum limit $|\omega_m|,v_Fq\ll E_F$ is assumed. It is clear that a finite magnetic field along the $z$-direction introduces anisotropy in the bosonic self-energy turning it into a non-trivial function of both the in-plane momentum $q_\parallel$ and $q_z$. In the absence of the fermionic self-energy insertions, Eq.~\eqref{eq:Pi_eq} acquires the following form
\begin{equation}
    \begin{aligned}
\Pi(\omega_l,&q_\parallel,q_z)=\frac{g^2 m\omega_c}{4\pi^2}\sum_{n,\bar{n}}\int \hspace{-0.3em}dk_z\frac{n_F(\epsilon_{\bar{n}(k_z+q_z)})-n_F(\epsilon_{nk_z})}{i\omega_l+\epsilon_{nk_z}-\epsilon_{\bar{n}(k_z+q_z)}}\\
        &\times e^{-\frac{q_\parallel^2l_B^2}{2}}\frac{n!}{\bar{n}!}\left(\frac{q_\parallel^2l_B^2}{2}\right)^{\bar{n}-n}\left[L_{n}^{\bar{n}-n}\left(\frac{q_\parallel^2l_B^2}{2}\right)\right]^2,
    \end{aligned}\label{eq:PI2}
\end{equation}
where $L_n^{\bar{n}}(x)$ is the associated Laguerre polynomial.

First, we consider the static limit of Eq.~\eqref{eq:PI2}, describing the oscillations of the compressibility. To this end, we set $\omega_m=0$ and $q_\parallel=0$ before taking $q_z\to0$. In this limit, $\Pi(\omega_m=0,q_\parallel=0,q_z)$ acquires the following form
\begin{equation}
\Pi(0,0,q_z)=\frac{g^2m\omega_c}{2\pi}\sum_{n}\int\frac{dk_z}{2\pi} \frac{n_F(\epsilon_{n(k_z+q_z)})-n_F(\epsilon_{nk_z})}{\epsilon_{nk_z}-\epsilon_{n(k_z+q_z)}},
\end{equation}
where we have used $ \int_0^\infty dx e^{-x}L_n(x)L_m(x)=\delta_{m,n}$. In the zero temperature limit, and also assuming $q_z\to0$, we obtain 
\begin{equation}
    \Pi(0,0,q_z\to 0)=\frac{g^2m\omega_c}{2\pi}\sum_{n}\int_{-\infty}^\infty \frac{dk_z}{2\pi} \delta(\epsilon_{nk_z}-\mu).
\end{equation}
After performing the Poisson summation and integrating over $k_z$ in the stationary phase approximation, we find
\begin{equation}
   \Pi(0,0,q_z\to 0)\approx g^2\nu\left(1+\frac{\omega_c}{4\mu}\right)+g^2\nu_{\text{ocs}}.
\end{equation}
Here the second term corresponds to a small correction to the smooth part of the compressibility, whereas the last term is given by Eq.~\eqref{eq:compress22} and contains oscillations.

Our next task is to evaluate Eq.~\eqref{eq:PI2} in the zero temperature limit and at the leading order in $\omega_c/\mu$, but now assuming that the external momenta are in the range $R_c^{-1}\ll q_\parallel,q_z \ll k_F$. At $T=0$, the Fermi distribution $n_F(x)$ reduces to a step function, so it is convenient to define an integer-valued function $N_c(k_z)$ (i.e. the number of filled Landau levels for a given $k_z$), such that 
\begin{equation}
\epsilon_{nk_z}>0~\text{for}~n>N_c(k_z), ~~~ \epsilon_{nk_z}<0~\text{for}~n<N_c(k_z).
\end{equation}
In the weak magnetic field  limit, and assuming that $k_z \ll k_F$ (the case when $k_z$ is close to $k_F$ requires special care, and we will return to it later), we have $N_c(k_z)\gg 1$, and thus, it can be approximated as $ N_c(k_z)\approx (\mu-k_z^2/2m)/\omega_c$. In these terms, the sum in \eqref{eq:PI2} can be compactly written as 
\begin{equation}
   \begin{aligned}
&\Pi(\omega_l,q_\parallel,q_z)=\frac{g^2m}{\pi}\int_{-k_F}^{k_F} \frac{dk_z}{2\pi} {\sum_n}'{\sum_{\bar{n}}}'\frac{n!}{\bar{n}!}\;e^{-\frac{q_\parallel^2l_B^2}{2}}\\
        &\frac{\bar{n}-n+\frac{k_zq_z}{m\omega_c}}{(\omega_l/\omega_c)^2+(\bar{n}-n+\frac{k_zq_z}{m\omega_c})^2}\left(\frac{q_\parallel^2l_B^2}{2}\right)^{\bar{n}-n}\left[L_{n}^{\bar{n}-n}\left(\frac{q_\parallel^2l_B^2}{2}\right)\right]^2
   \end{aligned}
\end{equation}
where ${\sum_n}'$ stands for the sum over $n$ with the constraint $n<N_c(k_z)$, whereas ${\sum_{\bar{n}}}'$ denotes the sum over $m$ with the constraint $\bar{n}>N_c(k_z+q_z)$. 
In the general case, we need to know the difference between the two summation bounds. Using the expression for $N_c(k_z)$, we obtain
\begin{equation}
    N_c(k_z)-N_c(k_z+q_z)=\frac{k_z q_z}{m_z\omega_c}.
\end{equation}
This indicates that the bound difference depends on the sign of $k_zq_z$. To proceed, we assume that $q_z>0$, and split the integral over $k_z$ into two terms: $\int_0^{k_F}...$ and $\int_{-k_F}^0...$, and change the variables $k_z\to-k_z$ in the second term. Then both $k_z$ and $q_z$ are positive, and we define 
\begin{equation}
    N_0=\frac{k_z q_z}{m_z\omega_c}>0, ~~~ t=\frac{q_\parallel^2\l_B^2}{2}.
\end{equation}
\begin{widetext}
 We then obtain
\begin{equation}
    \begin{aligned}
        \Pi(\omega_l,q_\parallel,q_z)=\frac{g^2m}{\pi}\int_0^{k_F} \frac{dk_z}{2\pi}&\left\{\sum_{s=1}^{N_0}\sum_{n=N_c-N_0}^{N_c-s}\frac{-s+N_0}{(\omega_l/\omega_c)^2+(s-N_0)^2}\frac{n!}{(n+s)!}t^se^{-t}[L_n^s(t)]^2\right.\\
        &+\sum_{s=0}^\infty\sum_{n=\max\{N_c-N_0-s,0\}}^{N_c}\frac{s+N_0}{(\omega_l/\omega_c)^2+(s+N_0)^2}\frac{n!}{(n+s)!}t^se^{-t}[L_n^s(t)]^2\\
    &+\left.\sum_{s=N_0}^\infty\sum_{n=\max\{N_c+N_0-s,0\}}^{N_c}\frac{s-N_0}{(\omega_l/\omega_c)^2+(s-N_0)^2}\frac{n!}{(n+s)!}t^se^{-t}[L_n^s(t)]^2\right\}.\label{eq:sn2}
    \end{aligned}
\end{equation}
where we have used the relation $L_n^{-s}(t)=\frac{(n-s)!}{n!}(-t)^sL_{n-s}^s(t)$.
Note that both $N_c$ and $N_0$ depend on $k_z$, thus the summation has to be performed before the integration over $k_z$.

We are interested in the case when the magnetic field is weak. In this case, typical $n$ is of the order of $N_c$ which is very large. On the other hand, we can keep $t$ finite, and arrive at the situation when $t\ll n\sim N_c$. In this limit, after approximating $L_n^s(t)$ using the Bessel function $J_s(t)$ and summing over $n$, we obtain
\begin{equation}
    \begin{aligned}
        \Pi(\omega_l,q_\parallel,q_z)
        &=\frac{g^2m}{\pi}\int_0^{k_F} \frac{dk_z}{2\pi}\left[1-\left(\frac{\omega_l}{\omega_c}\right)^2\int_0^\pi \frac{dy}{\pi}J_0\left(4\sqrt{N_ct}\sin\frac{y}{2}\right)\sum_{s=-\infty}^\infty\frac{\cos(sy)}{(\omega_l/\omega_c)^2+(s+N_0)^2}\right]
    \end{aligned}\label{eq:C13}
\end{equation}
Note for the $s$-summation we have used the following identities
\begin{equation}
    \begin{aligned}
        &\sum_{s=-\infty}^\infty J_s^2\left(2\sqrt{N_ct}\right)= 1, ~~~ J_s^2(x)=\int_0^\pi\frac{dy}{\pi}\cos(sy)J_0\left(2x\sin\frac{y}{2}\right)
    \end{aligned}
\end{equation}

The last term in \eqref{eq:C13} containing $s$-summation can be done using Poisson summation formula. Defining $\kappa N_c=\omega_l/\omega_c$, we have (for $N_0$ being integer valued)
\begin{equation}
    \begin{aligned}
        \sum_{s=-\infty}^\infty\frac{\cos(sy)}{(\kappa N_c)^2+(s+N_0)^2}=\frac{\pi}{\kappa N_c}\frac{\cosh[\kappa N_c(y-\pi)]}{\sinh\pi\kappa N_c}\cos(N_0 y)
    \end{aligned}
\end{equation}
Note the additional oscillation term $\cos(N_0y)$ is absent in 2D. To summarize, we have 
\begin{equation}
    \begin{aligned}
        \Pi(\omega_l,q_\parallel,q_z)
        &=\frac{g^2m}{\pi}\int_0^{k_F} \frac{dk_z}{2\pi}\left[1-\frac{\omega_l}{\omega_c} \Upsilon(N_0,N_c)\right],\\
        \Upsilon(N_0,N_c)&=\int_0^\pi dy J_0\left(4\sqrt{N_ct}\sin\frac{y}{2}\right)\frac{\cosh[\kappa N_c(y-\pi)]}{\sinh\pi\kappa N_c}\cos(N_0 y).
    \end{aligned}
\end{equation}
In the limit of small $\kappa$, $N_c\gg 1$, and $N_0\ll N_c$, we can expand the integrand near $y=0$, and obtain
\begin{equation}
    \begin{aligned}
        J_0\left(4\sqrt{N_ct}\sin\frac{y}{2}\right)&\approx J_0\left(2\sqrt{N_ct}y\right)+\frac{1}{12}\sqrt{N_ct}y^3J_1\left(2\sqrt{N_ct}y\right),\\
        \frac{\cosh[\kappa N_c(y-\pi)]}{\sinh\pi\kappa N_c}&\approx e^{-\kappa N_c y},~~~
        \cos(N_0 y)\approx 1-\frac{N_0^2y^2}{2}.
    \end{aligned}
 \end{equation} 
The contribution to $ \Upsilon(N_0,N_c)$ originating from the vicinity of the point $y=0$ (denoted as $\Upsilon_{0}$) is given by
\begin{equation}
    \begin{aligned}
         \Upsilon_{0}&\approx\int_0^\infty dy e^{-\kappa N_c y} J_0\left(2\sqrt{N_ct}y\right)+\frac{1}{12}\sqrt{N_ct}\int_0^\infty dy e^{-\kappa N_c y}y^3J_1\left(2\sqrt{N_ct}y\right)\\
        &-\frac{N_0^2}{2}\int_0^\infty dy e^{-\kappa N_c y} y^2 J_0\left(2\sqrt{N_ct}y\right)-\frac{N_0^2\sqrt{N_ct}}{24}\int_0^\infty dy e^{-\kappa N_c y}y^5J_1\left(2\sqrt{N_ct}y\right)\\
        =\frac{1}{\sqrt{4N_ct+(\kappa N_c)^2}}&+N_0^2\frac{2N_ct-(\kappa N_c)^2}{(4N_ct+(\kappa N_c)^2)^{5/2}}+\frac{2N_ct[(\kappa N_c)^2-N_c t]}{(4N_c t+(\kappa N_c)^2)^{7/2}}-30N_0^2N_ct\frac{2(N_ct)^2-6N_ct(\kappa N_c)^2+(\kappa N_c)^4}{(4N_ct+(\kappa N_c)^2)^{11/2}}
    \end{aligned}
\end{equation}  
\end{widetext}
Next, we need to evaluate the integral over $k_z$. In order to properly determine the UV cutoff we need to recall that our analysis relied on the asymptotic approximation for $L_n^s(t)$ which applies when $N_c(k_z)$ is large. However, this condition cannot be satisfied when $k_z\to k_F$, leading to the unphysical divergence. To resolve this issue, we impose a UV cutoff $\Lambda$ defined by the following condition: $N_c(\Lambda)\sim t$. This leads to $\Lambda\sim \sqrt{k_F^2-q_\parallel^2}$. It is also easy to check that the condition $N_c(k_z)\gg N_0$ is obeyed as long as $q_z\lesssim q_\parallel^2/(2k_F)$ and $k_z\ll \Lambda$. After performing the remaining integral over $k_z$ with the cutoff $\Lambda$, we find 
\begin{equation}
    \begin{aligned}
    \int_0^{\Lambda}dk_z\frac{\omega_l}{k_F\omega_c}\Upsilon_{0}
        \approx&\frac{\omega_l}{v_F \sqrt{q_\parallel^2+q_z^2}}\left(\frac{\pi}{2}-\frac{(eB)^2}{8k_Fq_\parallel^3}\right),\label{eq:int3}
    \end{aligned}
\end{equation}
where we retained only the leading order correction to the usual Landau damping. Next, we extract the oscillating contribution from the vicinity of the point $y=\pi$, which we denote as $\Upsilon_\pi$
\begin{equation}
    \begin{aligned}
        \Upsilon_\pi&=\int_0^\pi dy J_0\left(4\sqrt{N_ct}\cos\frac{y}{2}\right)\frac{\cosh\kappa N_c y}{\sinh\pi\kappa N_c}\cos(N_0 y)(-1)^{N_0}\\
        &\approx\frac{\sqrt{2\pi}}{\sinh(\pi\kappa N_c)}\frac{\sin\left(4\sqrt{N_ct}\right)}{4\sqrt{N_ct}},
    \end{aligned}
\end{equation}
where we have used $ N_0\ll t $. 

Finally, after integrating over $k_z$ and combining the resulting expression with the contribution from $\Upsilon_0$, Eq.~\eqref{eq:int3}, we obtain  
\begin{equation}
    \begin{aligned}
        \Pi(\omega_m,q_\parallel, q_z)&=-\frac{\pi g^2\nu|\omega_m|}{2 v_F\sqrt{q_\parallel^2+q_z^2}}\left(1-\frac{(eB)^2}{4\pi k_Fq_\parallel^3}\right)\\
        &+\frac{g^2\nu|\omega_m|}{v_Fq_\parallel}\sqrt{\frac{\pi}{2}}\frac{\sin\left( 2R_cq_\parallel\right)}{\sinh(\pi\omega_m/\omega_c)}.
    \end{aligned}
\end{equation}

\bibliography{DHVA_V3.bib}
\end{document}